\documentclass[aps,prc,twocolumn,groupedaddress]{revtex4}

\usepackage{graphicx}
\usepackage{amssymb}
\usepackage{amsmath}
\begin{document}


\title{{\em Ab initio} many-body calculations of nucleon scattering on $^4$He, $^7$Li, $^7$Be, $^{12}$C and $^{16}$O}


\author{Petr Navr{\'a}til$^1$, Robert Roth$^2$, and Sofia Quaglioni$^1$}
\affiliation{$^1$Lawrence Livermore National Laboratory, P.O. Box 808, L-414, Livermore, CA 94551, USA\\
$^2$Institut f\"{u}r Kernphysik, Technische Universit\"{a}t Darmstadt, 64289 Darmstadt, Germany}

%
\date{\today}
\begin{abstract}
We combine a recently developed {\it ab initio} many-body approach capable of describing simultaneously both bound and scattering states, the {\it ab initio} NCSM/RGM, with an importance truncation scheme for the cluster eigenstate basis and demostrate its applicability to nuclei with mass numbers as high as 17. Using soft similarity renormalization group evolved chiral nucleon-nucleon interactions, we first calculate nucleon-$^4$He phase shifts, cross sections and analyzing power. Next, we investigate nucleon scattering on $^7$Li, $^7$Be, $^{12}$C and $^{16}$O in coupled-channel NCSM/RGM calculations that include low-lying excited states of these nuclei. We check the convergence of phase shifts with the basis size and study $A=8$, $13$, and $17$ bound and unbound states. Our calculations predict low-lying resonances in $^8$Li and $^8$B that have not been experimentally clearly identified yet. We are able to reproduce reasonably well the structure of the $A=13$  low lying states. However, we find that $A=17$ states cannot be described without an improved treatment of $^{16}$O one-particle-one-hole excitations and $\alpha$ clustering.   
\end{abstract}

\pacs{21.60.De, 25.10.+s, 27.10.+h, 27.20.+n}

\maketitle

\section{Introduction}
\label{introduction}
Nuclei are quantum many-body systems with both bound and unbound states. A realistic {\it ab initio} description of light nuclei with predictive power must have the capability to describe both classes of states within a unified framework. Over the past decade, significant progress has been made in our understanding of the properties of the bound states of light nuclei starting from realistic nucleon-nucleon ($NN$) interactions, see e.g. Ref.~\cite{benchmark} and references therein, and more recently also from $NN$ plus three-nucleon ($NNN$) interactions~\cite{Nogga00,GFMC,NO03}. The solution of the nuclear many-body problem becomes more complex when scattering or nuclear reactions are considered. For $A=3$ and 4 nucleon systems, the Faddeev~\cite{Witala01} and Faddeev-Yakubovsky~\cite{Lazauskas05} as well
as the hyperspherical harmonics (HH) \cite{Pisa} or the Alt, Grassberger and Sandhas (AGS) 
\cite{Deltuva} methods are applicable and successful. However, {\em ab initio} calculations for unbound states and scattering processes involving more than four nucleons in total are quite challenging. The first {\it ab initio} many-body neutron-$^4$He scattering calculations were performed within the Green's Function Monte Carlo method using the Argonne $NN$ potential and the Illinois $NNN$ interaction~\cite{GFMC_nHe4}.  Also, resonances in He isotopes were investigated within the coupled-cluster method using the Gamow basis~\cite{Ha07}.

In a new development, we have recently combined  the {\em ab initio} no-core shell model (NCSM)~\cite{NCSMC12} and the resonating-group method (RGM)~\cite{RGM,RGM1,RGM2,RGM3,Lovas98,Hofmann08}, into a new many-body approach~\cite{NCSMRGM,NCSMRGM_PRC} ({\em ab initio} NCSM/RGM) capable of treating bound and scattering states of light nuclei in a unified formalism, starting from fundamental inter-nucleon interactions. The NCSM is an {\em ab initio} approach to the microscopic calculation of ground and low-lying excited states of light nuclei with realistic two- and, in general, three-nucleon forces. The RGM is a microscopic cluster technique based on the use of $A$-nucleon Hamiltonians, with fully anti-symmetric many-body wave functions built assuming that the nucleons are grouped into clusters. Although most of its applications are based on the use of binary-cluster wave functions, the RGM can be formulated for three (and, in principle, even more) clusters in relative motion~\cite{RGM1}. The use of the harmonic oscillator (HO) basis in the NCSM results in an incorrect description of the wave-function asymptotic and a lack of coupling to the continuum. By combining the NCSM with the RGM, we complement the ability of the RGM to deal with scattering and reactions with the use of realistic interactions, and a consistent  {\em ab initio} description of the nucleon clusters, achieved via the NCSM. Presently the NCSM/RGM approach has been formulated for processes involving binary-cluster systems only. However, extensions of the approach to include three-body cluster channels are feasible, also in view of recent developments on the treatment of both three-body bound and continuum states (see, e.g., Refs.~\cite{3bbound1,3bcont1,3bcnfr,3bbound2,3bcont2}). As described in detail in Refs.~\cite{NCSMRGM,NCSMRGM_PRC}, the {\em ab initio} NCSM/RGM approach has been already applied to study the $n\,$-${}^3$H, $n\,$-${}^4$He, $n\,$-${}^{10}$Be, and $p\,$-${}^{3,4}$He scattering processes,  and address the parity inversion  of  the $^{11}$Be  ground state, using realistic $NN$ potentials. In that work, we demonstrated convergence of the approach with increasing basis size in the case of the $A=4$ and $A=5$ scattering. The $n\,$-${}^{10}$Be calculations were, on the other hand, perfomed only in a limited basis due to the computational complexity of the NCSM calculations of the $^{10}$Be eigenstates. 

It is the purpose of the present paper to expand the applicability of the NCSM/RGM beyond the lightest nuclei by using sufficiently large $N\hbar\Omega$ HO excitations to guarantee convergence of the calculation with the HO basis expansion of both the cluster wave functions and the localized RGM integration kernels. The use of large $N\hbar\Omega$ values is now feasible due to the recent introduction of the importance truncated (IT) NCSM scheme~\cite{IT-NCSM,Roth09}. It turns out that many of the basis states used in the NCSM calculations are irrelevant for the description of any particular eigenstate, e.g., the ground state or a set of low-lying states. Therefore, if one were able to identify the important basis states beforehand, one could reduce the dimension of the matrix eigenvalue problem without losing predictive power. This can be done using an importance truncation scheme based on many-body perturbation theory~\cite{IT-NCSM}. 

We make use of the IT NCSM wave functions for the cluster eigenstates, in particular the eigenstates of the target nucleus of the binary nucleon-nucleus system, and calculate the one- and two-body densities that are then used to obtain the NCSM/RGM integration kernels. We benchmark the IT approach in basis sizes accessible by the full calculation and apply it within still larger basis sizes until convergence is reached for target nuclei as heavy as $^{12}$C or $^{16}$O. In this study, we employ a similarity renormalization group (SRG) ~\cite{SRG,Roth_SRG,Roth_PPNP} evolved chiral N$^3$LO $NN$ potential~\cite{N3LO} (SRG-N$^3$LO) that is soft enough to allow us reach convergence within about $14-16\hbar\Omega$ HO excitations in the basis expansion.

In Sect.~\ref{formalism}, we briefly overview the NCSM/RGM formalism and present for the first time the IT NCSM scheme that includes both ground and low-lying excited states in the set of reference states. Next, we present scattering calculation results for the $n$-$^4$He and $p$-$^4$He systems in Sect.~\ref{n4He}. In particular, we compare the calculated phase shifts to an R-matrix analysis of experimental data and, further, calculated differential cross sections and analyzing powers in the energy range ~6-19 MeV to the corresponding experimental  data. Neutron elastic and inelastic scattering on $^7$Li and proton elastic and inelastic scattering on $^7$Be are investigated in Sect.~\ref{n7Li}. We present phase shifts, cross sections and scattering lengths. We predict resonances in $^8$Li and $^8$Be that have not been clearly identified in experiments yet. In Sect.~\ref{n12C}, we discuss nucleon-$^{12}$C results for both bound and unbound states of $^{13}$C and $^{13}$N, obtained including two $^{12}$C bound states, the ground and the first $2^+$ state, in the NCSM/RGM coupled-channel calculations. In Sect.~\ref{n16O}, we present results for the nucleon-$^{16}$O system. In the NCSM/RGM coupled-channel calculations, we take into account the $^{16}$O ground state and up to the lowest three $^{16}$O negative-parity states. Conclusions are given in Sect.~\ref{conclusions}.

\section{Formalism}
\label{formalism}

\subsection{NCSM/RGM}
\label{NCSMRGM}

The {\it ab inito} NCSM/RGM approach was introduced in Ref.~\cite{NCSMRGM} with details of the formalism given in Ref.~\cite{NCSMRGM_PRC}. Here we give a brief overview of the main points. 

In the present paper, we limit ourselves to a two-cluster RGM, which is based on binary-cluster channel states of total angular momentum $J$, parity $\pi$, and isospin $T$,
\begin{eqnarray}
|\Phi^{J^\pi T}_{\nu r}\rangle &=& \Big [ \big ( \left|A{-}a\, \alpha_1 I_1^{\,\pi_1} T_1\right\rangle \left |a\,\alpha_2 I_2^{\,\pi_2} T_2\right\rangle\big ) ^{(s T)}\nonumber\\
&&\times\,Y_{\ell}\left(\hat r_{A-a,a}\right)\Big ]^{(J^\pi T)}\,\frac{\delta(r-r_{A-a,a})}{rr_{A-a,a}}\,.\label{basis}
\end{eqnarray}
In the above expression, $\left|A{-}a\, \alpha_1 I_1^{\,\pi_1} T_1\right\rangle$ and $\left |a\,\alpha_2 I_2^{\,\pi_2} T_2\right\rangle$ are the internal (antisymmetric) wave functions of the first and second cluster, containing $A{-}a$ and $a$ nucleons ($a{<}A$), respectively. They are characterized by angular momentum quantum numbers $I_1$ and $I_2$ coupled together to form channel spin $s$. For their parity, isospin and additional quantum numbers we use, respectively, the notations $\pi_i, T_i$, and $\alpha_i$, with $i=1,2$. The cluster centers of mass are separated by the relative coordinate 
\begin{equation}
\vec r_{A-a,a} = r_{A-a,a}\hat r_{A-a,a}= \frac{1}{A - a}\sum_{i = 1}^{A - a} \vec r_i - \frac{1}{a}\sum_{j = A - a + 1}^{A} \vec r_j\,,
\end{equation}
where $\{\vec{r}_i, i=1,2,\cdots,A\}$ are the $A$ single-particle coordinates.
The channel states~(\ref{basis}) have relative angular momentum $\ell$. It is convenient to group all relevant quantum numbers into a cumulative index $\nu=\{A{-}a\,\alpha_1I_1^{\,\pi_1} T_1;\, a\, \alpha_2 I_2^{\,\pi_2} T_2;\, s\ell\}$.
 
The former basis states can be used to expand the many-body wave function according to
\begin{equation}
|\Psi^{J^\pi T}\rangle = \sum_{\nu} \int dr \,r^2\frac{g^{J^\pi T}_\nu(r)}{r}\,\hat{\mathcal A}_{\nu}\,|\Phi^{J^\pi T}_{\nu r}\rangle\,. \label{trial}
\end{equation}
As the basis states~(\ref{basis}) are not anti-symmetric under exchange of nucleons belonging to different clusters, in order to preserve the Pauli principle one has to introduce the appropriate inter-cluster anti-symmetrizer, schematically
$\hat{\mathcal A}_{\nu}=\sqrt{\frac{(A{-}a)!a!}{A!}}\sum_{P}(-)^pP\,,$
where the sum runs over all possible permutations $P$ that can be carried out 
among nucleons pertaining to different clusters, and $p$ is the number of interchanges characterizing them. The coefficients of the expansion~(\ref{trial}) are the relative-motion wave functions $g^{J^\pi T}_\nu(r)$, which represent the only unknowns of the problem. To determine them one has to solve the non-local integro-differential coupled-channel equations 
\begin{equation}
\sum_{\nu}\int dr \,r^2\left[{\mathcal H}^{J^\pi T}_{\nu^\prime\nu}(r^\prime, r)-E\,{\mathcal N}^{J^\pi T}_{\nu^\prime\nu}(r^\prime,r)\right] \frac{g^{J^\pi T}_\nu(r)}{r} = 0\,,\label{RGMeq}
\end{equation}
where the two integration kernels, the Hamiltonian kernel,
\begin{equation}
{\mathcal H}^{J^\pi T}_{\nu^\prime\nu}(r^\prime, r) = \left\langle\Phi^{J^\pi T}_{\nu^\prime r^\prime}\right|\hat{\mathcal A}_{\nu^\prime}H\hat{\mathcal A}_{\nu}\left|\Phi^{J^\pi T}_{\nu r}\right\rangle\,,\label{H-kernel}
\end {equation}
and the norm kernel,
\begin{equation}
{\mathcal N}^{J^\pi T}_{\nu^\prime\nu}(r^\prime, r) = \left\langle\Phi^{J^\pi T}_{\nu^\prime r^\prime}\right|\hat{\mathcal A}_{\nu^\prime}\hat{\mathcal A}_{\nu}\left|\Phi^{J^\pi T}_{\nu r}\right\rangle\,,\label{N-kernel}
\end{equation}
contain all the nuclear structure and anti-symmetrization properties of the problem. In particular, the non-locality of the kernels is a direct consequence of the exchanges of nucleons between the clusters. We have used the notation $E$ and $H$ to denote the total energy in the center-of-mass frame, and the intrinsic $A$-nucleon microscopic Hamiltonian, respectively.
  
The formalism presented above is combined with the {\em ab initio} NCSM in two steps: 

First, we note that the Hamiltonian can be written as
\begin{equation}\label{Hamiltonian}
H=T_{\rm rel}(r)+ {\mathcal V}_{\rm rel} +\bar{V}_{\rm C}(r)+H_{(A-a)}+H_{(a)}\,,
\end{equation}
where $H_{(A-a)}$ and $H_{(a)}$ are the ($A{-}a$)- and $a$-nucleon intrinsic Hamiltonians, respectively, $T_{\rm rel}(r)$ is the relative kinetic energy 
and ${\mathcal V}_{\rm rel}$ is the sum of all interactions between nucleons belonging to different clusters after subtraction of the average Coulomb interaction between them, explicitly singled out in the term $\bar{V}_{\rm C}(r)=Z_{1\nu}Z_{2\nu}e^2/r$ ($Z_{1\nu}$ and $Z_{2\nu}$ being the charge numbers of the clusters in channel $\nu$).   We use identical realistic potentials in both the cluster's Hamiltonians and inter-cluster interaction ${\mathcal V}_{\rm rel}$. Accordingly, $\left|A{-}a\, \alpha_1 I_1^{\,\pi_1} T_1\right\rangle$ and $\left |a\,\alpha_2 I_2^{\,\pi_2} T_2\right\rangle$ are obtained by diagonalizing $H_{(A-a)}$ and $H_{(a)}$, respectively, in the model space spanned by the NCSM $N_{\rm max}\hbar\Omega$ HO basis. Note that in the present paper we use soft SRG evolved $NN$ potentials. Therefore, there is no need to derive any further effective interaction tailored to the model space truncation as with these soft interactions our calculations converge in the model spaces we are able to reach.

Second, we replace the delta functions in the localized parts of the Hamiltonian~(\ref{H-kernel}) and the norm~(\ref{N-kernel}) kernels with their representation in the HO model space. We use identical HO frequency as for the cluster eigenstate wave functions and a consistent model space size ($N_{\rm max}$). We emphasize that this replacement is performed only for the localized parts of the kernels. The diagonal parts coming from the identity operator in the antisymmetrizers, the kinetic term and the average Coulomb potential are treated exactly.

\begin{figure}
\begin{minipage}{6cm}
\rotatebox{-90}{ \includegraphics[height=.20\textheight]{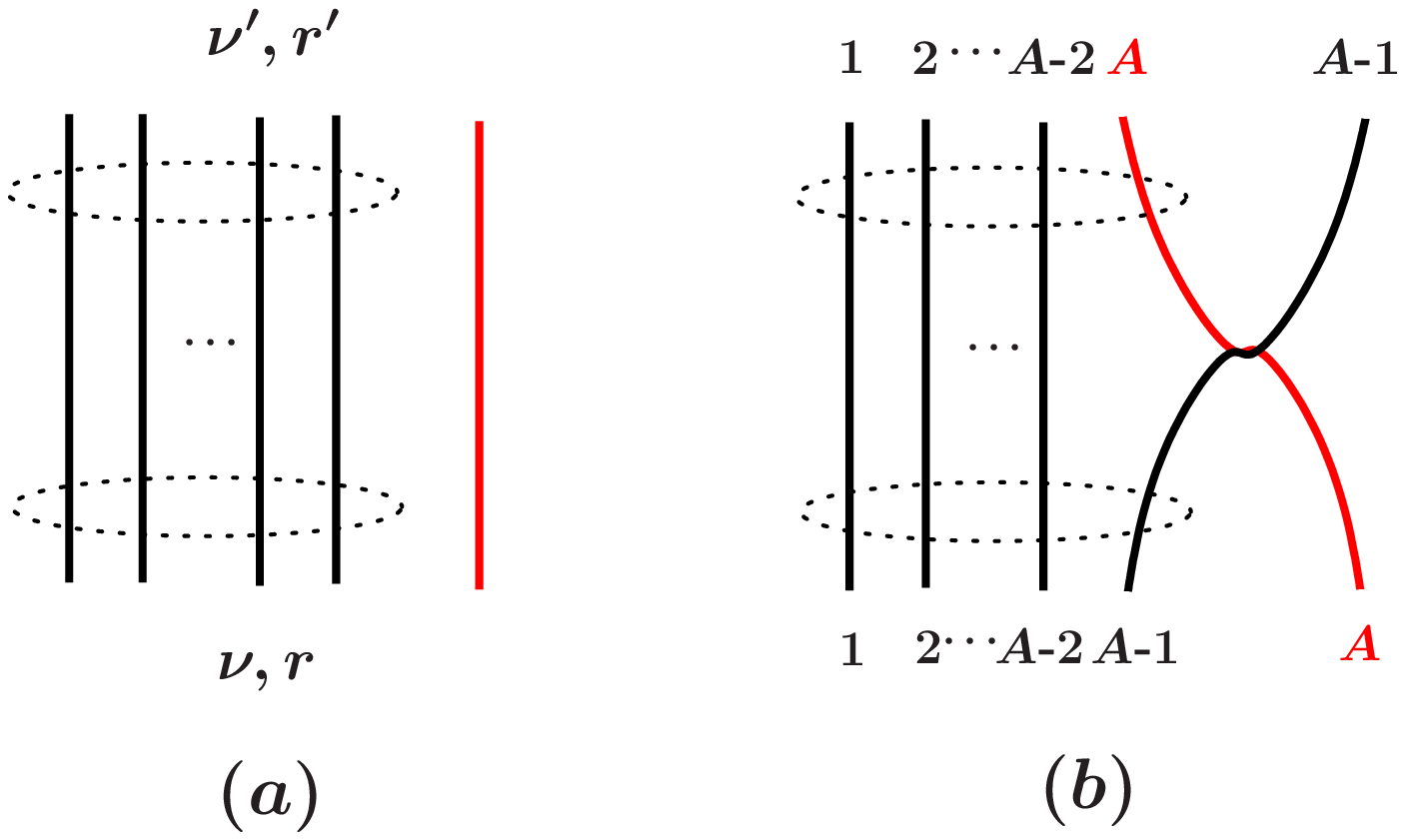}}
\end{minipage}
\begin{minipage}{8cm}
\rotatebox{-90}{\includegraphics[height=.35\textheight]{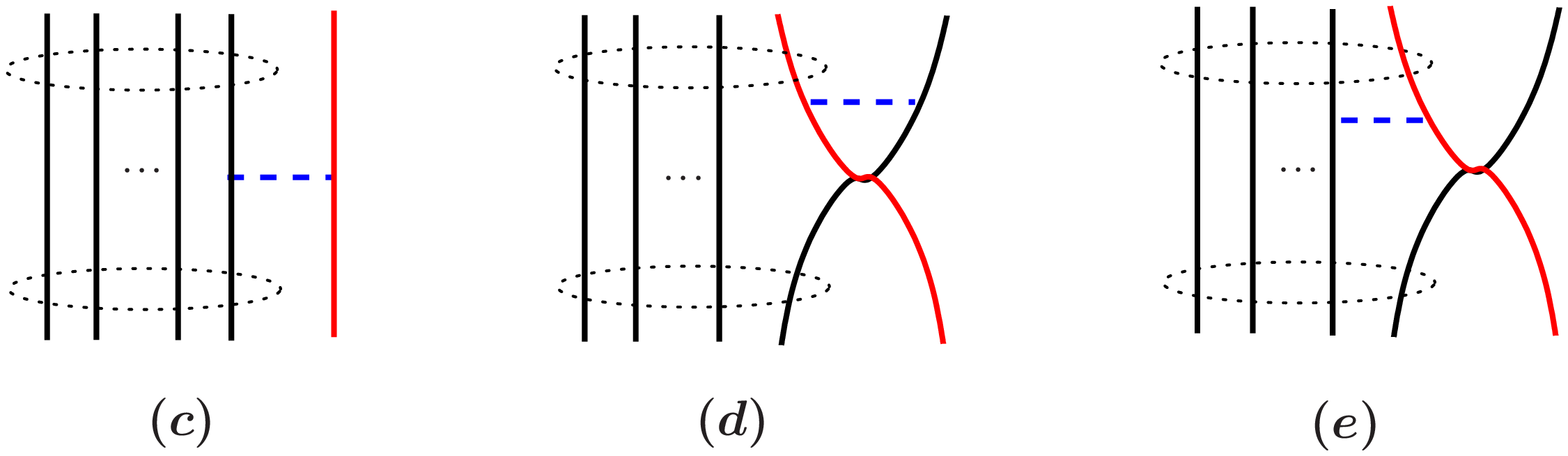}}
\end{minipage}
\caption{Diagrammatic representation of: ($a$) ``direct" and ($b$) ``exchange"  components of the norm kernel; ($c$ and $d$) ``direct"  and ($e$) ``exchange"   components of the potential kernel. The group of circled black lines represents the target cluster, a state of $A{-}1$ nucleons. The separate red line represents the projectile, a single nucleon. Bottom and upper part of the diagram represent initial and final states, respectively.}\label{diagram-norm-pot}
\end{figure}

In this paper, we apply the NCSM/RGM formalism in the single-nucleon projectile basis, i.e., for binary-cluster channel states~(\ref{basis}) with $a=1$ (with channel index $\nu = \{ A{-}1 \, \alpha_1 I_1^{\pi_1} T_1; \, 1\, \frac 1 2 \frac 1 2;\, s\ell\}$). As an illustration, let's discuss in more detail the norm kernel that is rather simple in this basis:
\begin{eqnarray}
{\mathcal N}^{J^\pi T}_{\nu^\prime\nu}(r^\prime, r)& = &\left\langle\Phi^{J^\pi T}_{\nu^\prime r^\prime}\right|1-\sum_{i=1}^{A-1}\hat P_{iA} \left|\Phi^{J^\pi T}_{\nu r}\right\rangle
\\
&=&\delta_{\nu^\prime\,\nu}\,\frac{\delta(r^\prime-r)}{r^\prime\,r}-(A-1)\sum_{n^\prime n}R_{n^\prime\ell^\prime}(r^\prime) R_{n\ell}(r)\nonumber\\
&&\times \left\langle\Phi^{J^\pi T}_{\nu^\prime n^\prime}\right|\hat P_{A-1,A} \left|\Phi^{J^\pi T}_{\nu n}\right\rangle\,.\label{norm}
\end{eqnarray}
We can easily recognize a direct term, in which initial and final state are identical (corresponding to diagram $(a)$ of Fig.~\ref{diagram-norm-pot}), and a many-body correction due to the exchange part of the inter-cluster anti-symmetrizer (corresponding to diagram $(b)$ of Fig.~\ref{diagram-norm-pot}). We note that in calculating the matrix elements of the exchange operator $\hat P_{A-1,A}$ we replaced the delta function of Eq.~(\ref{basis}) with its representation in the HO model space as discussed above. This is appropriate as the transposition $\hat P_{A-1,A}$ operator acting on the traget wave function is short-to-medium range. On the contrary, the $\delta$-function coming from the identity is treated exactly. The presence of the inter-cluster anti-symmetrizer affects also the Hamiltonian kernel, and, in particular, the matrix elements of the interaction. For a $NN$ potential one obtains a direct term involving interaction and exchange of two nucleons only (see diagrams ($c$) and ($d$) of Fig.~\ref{diagram-norm-pot}), and an exchange term involving three-nucleons. Diagram ($e$) of Fig.~\ref{diagram-norm-pot} describes this latter term, in which  the last nucleon is exchanged with one of the nucleons of the first cluster, and interacts with yet another nucleon. For more details on the integration kernels in the single-nucleon projectile basis we refer the readers to Ref.~\cite{NCSMRGM_PRC}.  

Being translationally-invariant  quantities, the norm and Hamiltonian kernels can be ``naturally" derived working within the NCSM Jacobi-coordinate basis. However, by introducing Slater-determinant channel states of the type 
\begin{eqnarray}
|\Phi^{J^\pi T}_{\nu n}\rangle_{\rm SD}   &=&    \Big [\big (\left|A{-}a\, \alpha_1 I_1 T_1\right\rangle_{\rm SD} 
\left |a\,\alpha_2 I_2 T_2\right\rangle\big )^{(s T)}\nonumber\\
&&\times Y_{\ell}(\hat R^{(a)}_{\rm c.m.})\Big ]^{(J^\pi T)} R_{n\ell}(R^{(a)}_{\rm c.m.})\,,
\label{SD-basis}
\end{eqnarray}
in which the eigenstates of the $(A{-}a)$-nucleon fragment are obtained in the SD basis (while the second cluster is still a NCSM Jacobi-coordinate eigenstate), it can be easily demonstrated that translationally invariant matrix elements can be extracted from those calculated in the SD basis of Eq.~(\ref{SD-basis}) by inverting the following expression:
 \begin{eqnarray}
&& {}_{\rm SD}\!\left\langle\Phi^{J^\pi T}_{\nu^\prime n^\prime}\right|\hat{\mathcal O}_{\rm t.i.}\left|\Phi^{J^\pi T}_{\nu n}\right\rangle\!{}_{\rm SD} = \nonumber\\
&&\nonumber\\
&&\sum_{n^\prime_r \ell^\prime_r, n_r\ell_r, J_r}
 \left\langle\Phi^{J_r^{\pi_r} T}_{\nu^\prime_r n^\prime_r}\right|\hat{\mathcal O}_{\rm t.i.}\left|\Phi^{J_r^{\pi_r} T}_{\nu_r n_r}\right\rangle\nonumber\\
&&  \times \sum_{NL} \hat \ell \hat \ell^\prime \hat J_r^2 (-1)^{(s+\ell-s^\prime-\ell^\prime)}
  \left\{\begin{array}{ccc}
 s &\ell_r&  J_r\\
  L& J & \ell
 \end{array}\right\}
 \left\{\begin{array}{ccc}
 s^\prime &\ell^\prime_r&  J_r\\
  L& J & \ell^\prime
 \end{array}\right\}\nonumber\\
 &&\nonumber\\
&& \times\langle  n_r\ell_rNL\ell | 00n\ell\ell \rangle_{\frac{a}{A-a}} 
 \;\langle  n^\prime_r\ell^\prime_rNL\ell | 00n^\prime\ell^\prime\ell^\prime \rangle_{\frac{a}{A-a}} \,.\label{Oti}
 \end{eqnarray}
Here $\hat {\mathcal O}_{\rm t.i.}$ represents any scalar and parity-conserving translational-invariant operator ($\hat {\mathcal O}_{\rm t.i.} = \hat{\mathcal A}$, $\hat{\mathcal A} H \hat{\mathcal A}$, etc.).
We exploited both Jacobi-coordinate and SD channel states to verify our results.  The use of the SD basis is computationally advantageous and allows us to explore reactions involving $p$-shell nuclei, as done in the present work. In order to calculate the parts of the integration kernels depicted in Fig.~\ref{diagram-norm-pot}  (b), (c) and (d), all information that we need from the SD basis calculation are one-body densities of the target eigenstates. For the (e) part of the integration kernel in Fig.~\ref{diagram-norm-pot}, we need two-body densities of the target eigenstates obtained in the SD basis.

Due to the presence of the norm kernel ${\mathcal N}^{J^\pi T}_{\nu^\prime\nu}(r^\prime, r)$, Eq.~(\ref{RGMeq}) does not represent a system of multichannel Schr\"odinger equations, and $g^{J^\pi T}_\nu(r)$ do not represent Schr\"odinger wave functions. The short-range non-orthogonality, induced by the non-identical permutations in the inter-cluster anti-symmetrizers, can be removed by introducing normalized Schr\"odinger wave functions
\begin{equation}
\frac{\chi^{J^\pi T}_\nu(r)}{r} = \sum_{\gamma}\int dy\, y^2 {\mathcal N}^{\frac12}_{\nu\gamma}(r,y)\,\frac{g^{J^\pi T}_\gamma(y)}{y}\,,
\end{equation}
where ${\mathcal N}^{\frac12}$ is the square root of the norm kernel, and applying the inverse-square root of the norm kernel, ${\mathcal N}^{-\frac12}$, to both left and right-hand side of the square brackets in Eq.~(\ref{RGMeq}).  This procedure, explained in more detail in Ref.~\cite{NCSMRGM_PRC}, leads to a system of multichannel Schr\"odinger equations
\begin{eqnarray}
&&[\hat T_{\rm rel}(r) + \bar V_{\rm C}(r) -(E - E_{\alpha_1}^{I_1^{\pi_1} T_1} - E_{\alpha_2}^{I_2^{\pi_2} T_2})]\frac{\chi^{J^\pi T}_{\nu} (r)}{r} \nonumber\\[2mm]
&&+ \sum_{\nu^\prime}\int dr^\prime\,r^{\prime\,2} \,W^{J^\pi T}_{\nu \nu^\prime}(r,r^\prime)\,\frac{\chi^{J^\pi T}_{\nu^\prime}(r^\prime)}{r^\prime} = 0,\label{r-matrix-eq}
\end{eqnarray} 
where $E_{\alpha_i}^{I_i^{\pi_i} T_i}$ is the energy eigenvalue of the $i$-th cluster ($i=1,2$), and $W^{J^\pi T}_{\nu^\prime \nu}(r^\prime,r)$ is the overall non-local potential between the two clusters, which depends on the channel of relative motion, while it does not depend on the energy. These are the equations that we finally solve to obtain both our scattering and bound-state results.

\subsection{Importance truncated NCSM with excited states}
\label{ITNCSM}

The primary limitation for the range of applicability of the NCSM in terms of particle number $A$ and model spaces size $N_{\max}$ results from the factorial growth of the dimension of the $N_{\max}\hbar\Omega$ space. Except for light isotopes, it is hardly possible to obtain a converged result using a 'bare' Hamiltonian within the $N_{\max}\hbar\Omega$ spaces that are computationally tractable. 

At this point the importance truncation offers a solution. The importance truncation in connection with the NCSM was introduced in Ref.~\cite{IT-NCSM} and discussed in detail in Ref.~\cite{Roth09}. In the following we summarize a few key features of the IT-NCSM and generalize the approach to the simultaneous description of excited states.

The motivation for the importance truncation results from the observation that the expansion of any particular eigenstate of the Hamiltonian in a full $m$-scheme NCSM space typically contains a large number of basis states with extremely small or vanishing amplitudes. The amplitudes define an adaptive truncation criterion, which takes into account the properties of the Hamiltonian and the structure of the eigenstate under consideration. If those amplitudes were known---at least approximately---before actually solving the eigenvalue problem, one could reduce the model space to the most relevant basis states by imposing a threshold on the amplitude. The amplitude of a particular basis state $| \Phi_{\nu} \rangle$ in the expansion of a specific eigenstate can be estimated using first-order multiconfigurational perturbation theory. In order to set up a perturbation series we need an initial approximation of the target state, the so-called reference state $| \Psi_{\text{ref}} \rangle$. In practice this reference state will be a superposition of basis states $| \Phi_{\mu} \rangle \in \mathcal{M}_{\text{ref}}$ from a reference space $\mathcal{M}_{\text{ref}}$:
\begin{equation}
\label{eq:itncsm_referencestate}
  | \Psi_{\text{ref}} \rangle 
  = \sum_{\mu \in \mathcal{M}_{\text{ref}}} C_{\mu}^{(\text{ref})} | \Phi_{\mu} \rangle \;.
\end{equation}
The reference state and the amplitudes $C_{\mu}^{(\text{ref})}$ are typically extracted from a previous NCSM calculation. Based on $| \Psi_{\text{ref}} \rangle$ as unperturbed state, we can evaluate the first-order perturbative correction to the target state resulting from basis states $| \Phi_{\nu} \rangle \notin \mathcal{M}_{\text{ref}}$. Their first-order amplitude defines the so-called importance measure
\begin{equation}
\label{eq:itncsm_importancemeasure}
  \kappa_{\nu} 
  = -\frac{\langle \Phi_\nu | H | \Psi_{\text{ref}} \rangle}{\epsilon_\nu - \epsilon_{\text{ref}}} 
  = -\sum_{\mu\in\mathcal{M}_{\text{ref}}} C_{\mu}^{(\text{ref})} \frac{\langle \Phi_\nu | H | \Phi_\mu \rangle}{\epsilon_\nu - \epsilon_{\text{ref}}} \;.
\end{equation}
The energy denominator $\epsilon_\nu - \epsilon_{\text{ref}}$ in a M\o{}ller-Plesset-type partitioning is simply given by the unperturbed harmonic-oscillator excitation energy of the basis state $| \Phi_{\nu} \rangle$ (see Ref.~\cite{Roth09} for details). 

Imposing an importance threshold $\kappa_{\min}$, we construct an importance truncated model space including all basis states with importance measure $|\kappa_{\nu}| \geq \kappa_{\min}$. Since the importance measure is zero for all basis states that differ from all of the  states in $\mathcal{M}_{\text{ref}}$ by more than a two-particle-two-hole excitation, we have to embed the construction of the importance truncated space into an iterative update cycle. After constructing the importance truncated space and solving the eigenvalue problem in that space, we obtain an improved approximation for the target state that defines a reference state for the next iteration. In order to accelerate the evaluation of the importance measure \eqref{eq:itncsm_importancemeasure}, we typically do not  use the complete eigenstate as new reference state, but project it onto a reference space spanned by the basis states with $|C_{\nu}|\geq C_{\min}$, where $C_{\nu}$ are the coefficients resulting from the solution of the eigenvalue problem. The second threshold $C_{\min}$ will be chosen sufficiently small so as not to affect the results for a given $\kappa_{\min}$ threshold. 

Simple iterative update schemes can be devised for any type of full model spaces, as discussed in Refs.~\cite{Roth09,RoGo09}. Specifically for the $N_{\max}\hbar\Omega$ space of the NCSM, however, there is an efficient sequential update scheme leading to the IT-NCSM(seq) approach. It is based on the fact that all states of an $(N_{\max}+2)\hbar\Omega$ space can be generated from the basis states of an $N_{\max}\hbar\Omega$ space using two-particle-two-hole excitations at most. Thus a single importance update starting from a reference state in an $N_{\max}\hbar\Omega$ space gives access to all relevant states in an $(N_{\max}+2)\hbar\Omega$ space. Making use of this property, in the IT-NCSM(seq) we start with a full NCSM calculation in, e.g., $2\hbar\Omega$ and use this eigenstate after applying the $C_{\min}$ threshold as reference state for constructing the importance truncated $4\hbar\Omega$ space. After solving the eigenvalue problem for this importance truncated $4\hbar\Omega$ space we use the resulting eigenstate as reference state to construct the $6\hbar\Omega$ space, and so on. Thus only one importance update is required for each value of $N_{\max}$, which makes this scheme very efficient computationally. Moreover, in the limit of vanishing thresholds, $(\kappa_{\min},C_{\min})\to0$, this scheme recovers the full $N_{\max}\hbar\Omega$ space at each step of the sequence, i.e., the IT-NCSM(seq) would recover the full NCSM result.

Based on this limiting property, we can obtain a numerical approximation to the full NCSM result by extrapolating the IT-NCSM(seq) observables obtained for a set of different importance thresholds $\kappa_{\min}$ (and in principle also $C_{\min}$) to $\kappa_{\min}\to0$. Through this extrapolation, the contribution of discarded basis states, i.e. those with importance measures $|\kappa_{\nu}|$ below the smallest threshold considered, is effectively recovered. Because the control parameter $\kappa_{\min}$ is tied to the physical structure of the eigenstate, we observe a smooth threshold dependence for all observables, which allows for a robust threshold extrapolation. In the case of the energy we can improve the quality of the extrapolation further by considering a perturbative second-order estimate for the energy of the excluded basis states. While setting up the importance truncated space, all second-order energy contributions
\begin{equation}
\label{eq:itncsm_importancemeasureenergy}
  \xi_{\nu} 
  = -\frac{| \langle \Phi_\nu | H | \Psi_{\text{ref}} \rangle |^2}{\epsilon_\nu - \epsilon_{\text{ref}}} \;.
\end{equation}
for the discarded states with $|\kappa_{\nu}| < \kappa_{\min}$ are summed up to provide a correction $\Delta_{\text{excl}}(\kappa_{\min})$ to the energy eigenvalue. By construction this correction goes to zero in the limit $\kappa_{\min}\to0$. We use this additional information for a constrained simultaneous extrapolation of the energy to vanishing threshold with and without perturbative correction for the excluded states as described in detail in Ref.~\cite{Roth09}.

The whole concept can be generalized to the description of excited states. For the present application in connection with the NCSM/RGM, we aim at an importance truncated model space that is equally well suited for the description of the lowest $M$ eigenstates of the Hamiltonian for given parity and total angular momentum projection. Instead of using a single reference state, we employ different reference states $| \Psi_{\text{ref}}^{(m)} \rangle$, with $m=1,...,M$, for each of the $M$ target states. For each reference state we define a separate importance measure $\kappa_{\nu}^{(m)}$ following Eq. \eqref{eq:itncsm_importancemeasure}. A basis state $|\Phi_{\nu} \rangle$ is included in the importance truncated space if at least one of the importance measures $|\kappa_{\nu}^{(m)}|$ is above the threshold $\kappa_{\min}$, i.e., if it is relevant for the description of at least one of the $M$ target states it will be included. Because the different eigenstates are typically dominated by different basis states, the dimension of the importance truncated space grows linearly with $M$. 

In the IT-NCSM(seq) scheme we start with a full NCSM calculation in $2\hbar\Omega$ and use the lowest $M$ eigenstates as initial reference states $| \Psi_{\text{ref}}^{(m)} \rangle$. Based on the corresponding importance measures $\kappa_{\nu}^{(m)}$ the importance truncated $4\hbar\Omega$ space is constructed and the lowest $M$ eigenvectors in this space serve as new reference states (after application of the $C_{\min}$ threshold) for the construction of the $6\hbar\Omega$ space, and so on. From a sequence of IT-NCSM(seq) calculations we obtain a set of $M$ eigenvectors for each value of $N_{\max}$ which can be used to evaluate other observables.

By default we compute the expectation values of $\vec{J}^2$ and $\vec{T}^2$ as well as the expectation values of $H_{\text{int}}$ and $H_{\text{cm}}$. Indeed, since we use an importance truncated space in the $m$-scheme without explicit angular momentum projection, the eigenstates are not guaranteed to have good angular momentum and isospin. We therefore monitor the expectation values of $\vec{J}^2$ and $\vec{T}^2$ and find values which typically differ by less then $10^{-3}$ from the exact quantum numbers. As in the full NCSM we separate spurious center-of-mass (CM) excitations from the physical spectrum by adding a Lawson term $\beta H_{\text{cm}}$ to the translationally invariant intrinsic Hamiltonian $H_{\text{int}}$ (with the typical choice $\beta=10$). The use of this modified Hamiltonian provides at the same time a diagnostic for potential CM contaminations of the intrinsic states induced by the importance truncation. As discussed in Refs. \cite{RoGo09b,Roth09}, the independence of the intrinsic energies $\langle H_{\text{int}} \rangle$ on $\beta$ and the smallness of $\langle H_{\text{cm}} \rangle$ demonstate that the IT-NCSM(seq) solutions are free of CM contaminations. 

Eventually, the wave functions obtained in the IT-NCSM(seq) together with the threshold extrapolated intrinsic energies form the input for the NCSM/RGM calculations discussed in the following.

\section{Nucleon-$^4$He scattering}
\label{n4He}

The purpose of the nucleon-$^4$He calculations presented in this paper is two-fold. First, we want to check the predictive power of the SRG evolved chiral interaction in the $A=5$ system, where a lot of experimental scattering data exist and where our calculations can be easily converged with respect to the size of the basis expansion. Second, we want to benchmark the importance truncation scheme with the full-space calculations all the way up to very large $N_{\rm max}\hbar\Omega$ spaces. 

The first {\it ab initio} $A=5$ scattering calculations was reported in Ref.~\cite{GFMC_nHe4}. The $n$-$\alpha$ low-lying  $J^\pi=3/2^-$ and $1/2^-$ $P$-wave resonances as well as the $1/2^+$ $S$-wave non-resonant scattering below 5 MeV c.m.\ energy were obtained using the AV18 $NN$ potential with and without the three-nucleon force, chosen to be either the Urbana IX or the Illinois-2 model.  The results of these Green's function Monte Carlo (GFMC) calculations revealed sensitivity to the inter-nucleon interaction, and in particular to the strength of the spin-orbit force. 

Soon after, the development of the {\em ab initio} NCSM/RGM approach allowed us to calculate both $n$- and (for the first time) $p$-$\alpha$ scattering phase shifts for energies up to the inelastic threshold~\cite{NCSMRGM,NCSMRGM_PRC}, using several realistic $NN$ potentials, including the chiral N$^3$LO~\cite{N3LO}, the $V_{{\rm low}k}$~\cite{BoKu03} and the CD-Bonn~\cite{CD-Bonn2000} $NN$ potentials.  
Nucleon-$\alpha$ scattering provides one of the best-case scenarios for the application of the  NCSM/RGM approach.  This process is characterized by a single open channel up to the $d+^3$H threshold, which is fairly high in energy.   In addition, the low-lying resonances of the $^4$He nucleus are narrow enough to be reasonably reproduced diagonalizing the four-body Hamiltonian in the NCSM model space. 
In the present work we include the first excited state of $^4$He, the $0^+ 0$ state, as a closed channel in our NCSM/RGM basis space.

\subsection{Convergence with the size of the HO basis expansion}
\label{conv}

\begin{figure}[t]
\includegraphics*[width=1.0\columnwidth]{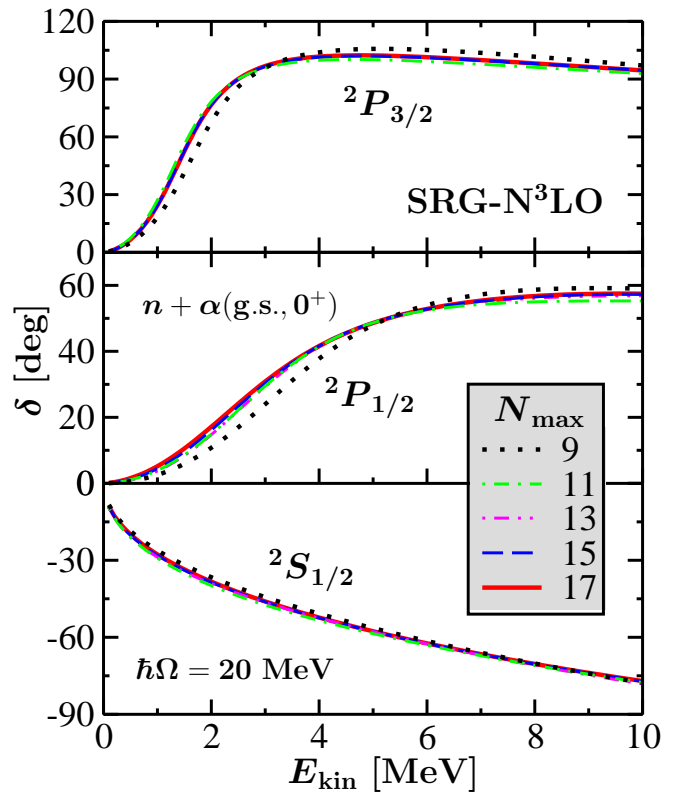}%
\caption{(Color online) Dependence of the $n$-$^4$He phase shifts on the size of the HO basis expansion of the $^4$He wave functions and the localized parts of the integration kernels. The $^4$He ground state and the first $0^+ 0$ excited states were included. The SRG-N$^3$LO $NN$ potential with a cutoff of 2.02 fm$^{-1}$ and the HO frequency $\hbar\Omega=20$ MeV were used.}
\label{fig:nHe4_phaseconv}
\end{figure}

We performed extensive nucleon-$^4$He calculations with the SRG-N$^3$LO $NN$ potential  with a cutoff of 2.02 fm$^{-1}$ to check convergence of our NCSM/RGM calculations. In Fig.~\ref{fig:nHe4_phaseconv}, we present  $n$-$^4$He phase shift results for the $S$- and $P$-waves obtained using an HO basis expansion up to $N_{\rm max}=17$ for  for the localized parts of the NCSM/RGM integration kernels and for the $^4$He ground- and the first-excited $0^+ 0$ wave functions (since these states have positive parity, the $N_{\rm max}-1$ expansion is in fact used for the $^4$He eigenstates). As seen in the figure, the phase-shift convergence is excellent. In particular, the $N_{\rm max}=17$ and the $N_{\rm max}=15$ curves lie on top of each other. The convergence rate demonstrated here is quite similar to that obtained using the $V_{{\rm low} k}$ $NN$ potentatial in our erlier study (compare the present Fig.~\ref{fig:nHe4_phaseconv} to the left panel of Fig. 13 in Ref.~\cite{NCSMRGM_PRC}). 

\subsection{Benchmark of Importance-Truncated calculations}
\label{IT_bench}

As shown in the previous subsection, for the $A=5$ system we are able to reach complete convergence with $^4$He wave functions obtained within full, non-truncated, NCSM calculations. We can, therefore, test the performace of the IT-NCSM scheme in this system all the way up to very large $N_{\rm max}$ values and see how well the IT-NCSM scheme reproduces the completetly converged results. It should be noted that for heavier $A=8,13$ and $A=17$ systems investigated later, full, non-truncated NCSM calculations for the $A=7$ ($A=12,16$) target nuclei are feasible only up to $N_{\rm max}=10$ ($N_{\rm max}=8$). It is, therefore, desirable and important to benchmark the IT-NCSM calculations in a lighter system like $A=5$ in  $N_{\rm max}>10$ calculations.

\begin{figure}[t]
\includegraphics*[width=1.0\columnwidth]{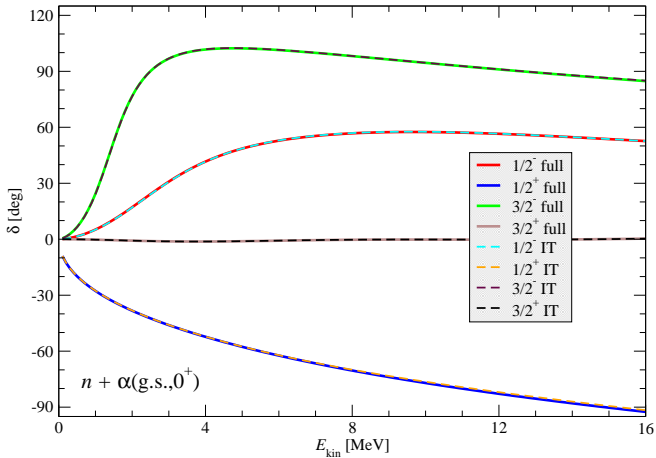}%
\caption{(Color online) Calculated $n$-$^4$He $S$- and $P$-wave phase shifts. Results obtained with $^4$He wave functions from full NCSM (solid lines) and IT-NCSM (dashed lines) calculations are compared. The SRG-N$^3$LO $NN$ potential with a cutoff of 2.02 fm$^{-1}$, the $N_{\rm max}=17$ basis space and the HO frequency $\hbar\Omega=20$ MeV were used. See text for details on the IT-NCSM calculation.}
\label{fig:nHe4_phase_full_IT}
\end{figure}

In Fig.~\ref{fig:nHe4_phase_full_IT}, we compare $n$-$^4$He phase shifts calculated within the NCSM/RGM with $^4$He wave functions obtained in a full $N_{\rm max}=16$ NCSM calculation and those obtained using $^4$He wave functions obtained within an $N_{\rm max}=16$  IT-NCSM calculation. The agreement of the two sets of phase shifts is excellent. It should be noted that the dimension of the full $N_{\rm max}=16$  $^4$He NCSM basis is 6344119. The dimension of the IT-NCSM basis used here to calculate the $^4$He wave functions was just 992578, more than a factor of six smaller.  Truncation parameters  $\kappa_{\min}=10^{-5} $  and $C_{\min}=2\times 10^{-4}$ were used. The ground state energy from the full NCSM calculation is $-28.224$ MeV. The $\kappa_{\min} \rightarrow 0$ extrapolated ground state energy from the IT-NCSM calculation is $-28.217(5)$ MeV with a difference from the full result less than 10 keV. The excited $0^+ 0$ energy obtained in the full NCSM calculation was 21.58 MeV. The corresponding extrapolated IT-NCSM result was 21.4(1) MeV. The slightly lower accuracy of the excited state reproduction in the IT-NCSM calculation is manifested in a very small deviation of the $S$-wave phase shift at energies above 12 MeV (less than 1 degree at 16 MeV). It should be noted that the excited $0^+ 0$ state is not bound. Consequently, it is challenging to reproduce the excited state as well as the ground state in a importance-truncated calculation. It should be also pointed out that unlike for the energies, no phase shift extrapolation was performed. The needed one- and two-body densities were calculated from the wave functions obtained in the IT-NCSM calculation with the truncation parameters described above. The excellent agreement of the full and the IT-NCSM phase shifts demonstrates that no extrapolation was actually necessary. Obviously, we can check the dependence of observables like phase shifts on the $\kappa_{\min}$ and  $C_{\min}$ and perform an extrapolation to vanishing values of these parameters if needed.

\subsection{Comparison with experimental data}
\label{n-He4_vs_exper}
\begin{figure}[t]
\includegraphics*[width=1.0\columnwidth]{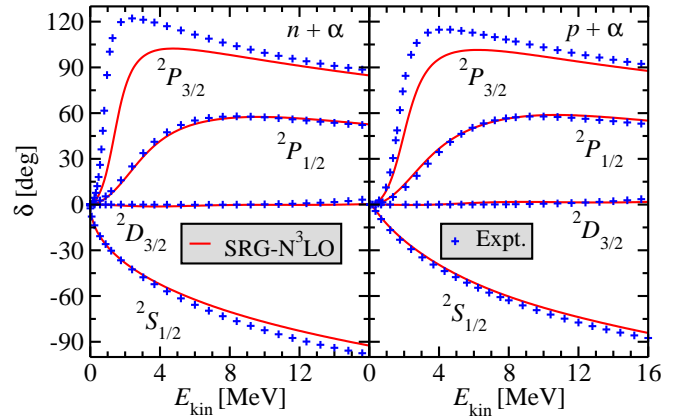}%
\caption{(Color online) Calculated $n-^4$He (left panels) and $p-^4$He (right panels) compared to the R-matrix analysis of experimental data~\cite{HalePriv}. The NCSM/RGM calculations that included the $^4$He ground state and the $0^+ 0$ excited state were done using the SRG-N$^3$LO $NN$ potential with a cutoff of 2.02 fm$^{-1}$. The HO frequency $\hbar\Omega=20$ MeV and $N_{\rm max}=17$ basis space were employed.}
\label{fig:N4He_phase}
\end{figure}
Our calculated $n$-$^4$He and $p$-$^4$He phase shifts are compared to those obtained from an $R$-matrix analysis of $N-^4$He experimental data~\cite{HalePriv} in Fig.~\ref{fig:N4He_phase}. The agreement is quite reasonable for the $S$-wave, $D$-wave and $^2P_{1/2}$-wave. The $^2P_{3/2}$ resonance is positioned at higher energy in the calculation and the corresponding phase shifts are underestimated with respect to the $R$-matrix results, although the disagreement becomes less and less pronounced starting at about 8 MeV. While the inclusion of negative-parity excited states of the $\alpha-$particle would likely increase somewhat  the $^2P_{3/2}$ phase shifts~\cite{NCSMRGM,NCSMRGM_PRC}, the observed difference is largely due to a reduction in spin-orbit strength caused by the neglect of the three-nucleon interaction in our calculations. The importance of the three-nucleon force in reproducing  the $R$-matrix $^2P_{3/2}$ phase shifts was demonstrated in the GFMC $n$-$^4$He calculations of Ref.~\cite{GFMC_nHe4}.  Overall, the present results obtained with the SRG-N$^3$LO $NN$ interaction agree better with experiment than our earlier calculations~\cite{NCSMRGM,NCSMRGM_PRC} with the $V_{{\rm low} k}$, N$^3$LO and CD-Bonn $NN$ potentials. The only exception is the $S$-wave phase shift which is best described using the CD-Bonn $NN$ potential. The larger spin-orbit strength of the employed SRG-N$^3$LO potential with respect to N$^3$LO itself is the likely responsible for the improved agreement. 

\begin{figure*}
\includegraphics*[width=1\columnwidth, angle = -90]{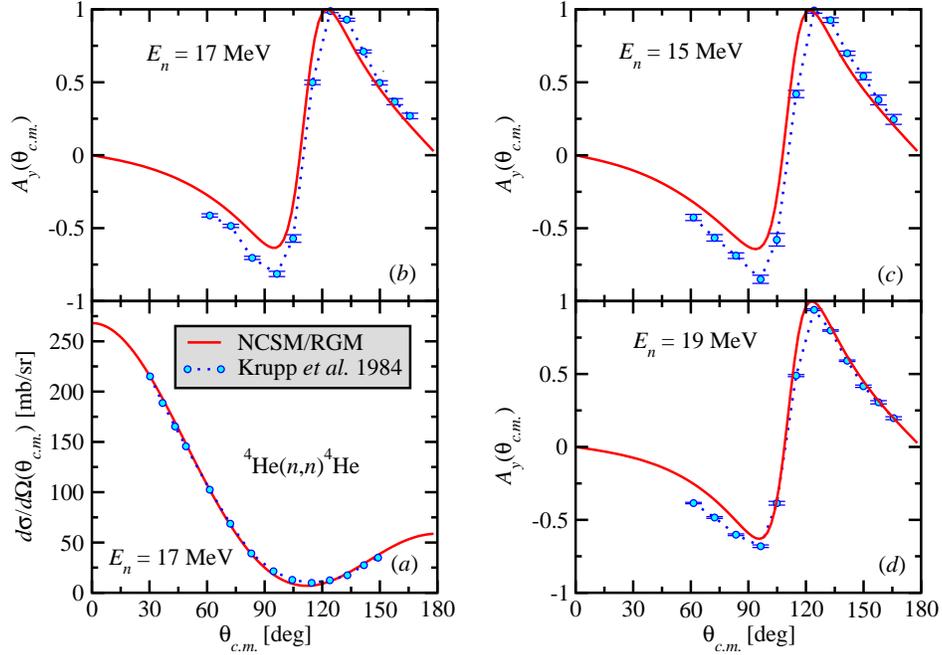}
\caption{(Color online) Calculated $n-^4$He differential cross section for neutron laboratory energy of $E_n = 17$ MeV ($a$), and analyzing power for $E_n=17$ ($b$), 15 ($c$) and 19 MeV ($d$) compared to experimental data from Ref.~\cite{Karlsruhe}. The NCSM/RGM results include the $^4$He ground state and first $0^+ 0$ excited state and were obtained using the SRG-N$^3$LO $NN$ potential with a cutoff of 2.02 fm$^{-1}$, for an HO frequency $\hbar\Omega=20$ MeV and basis space size $N_{\rm max}=17$.}
\label{fig:N4He_17MeVxsay}
\end{figure*}
\begin{figure*}
\includegraphics*[width=1.0\columnwidth, angle = -90]{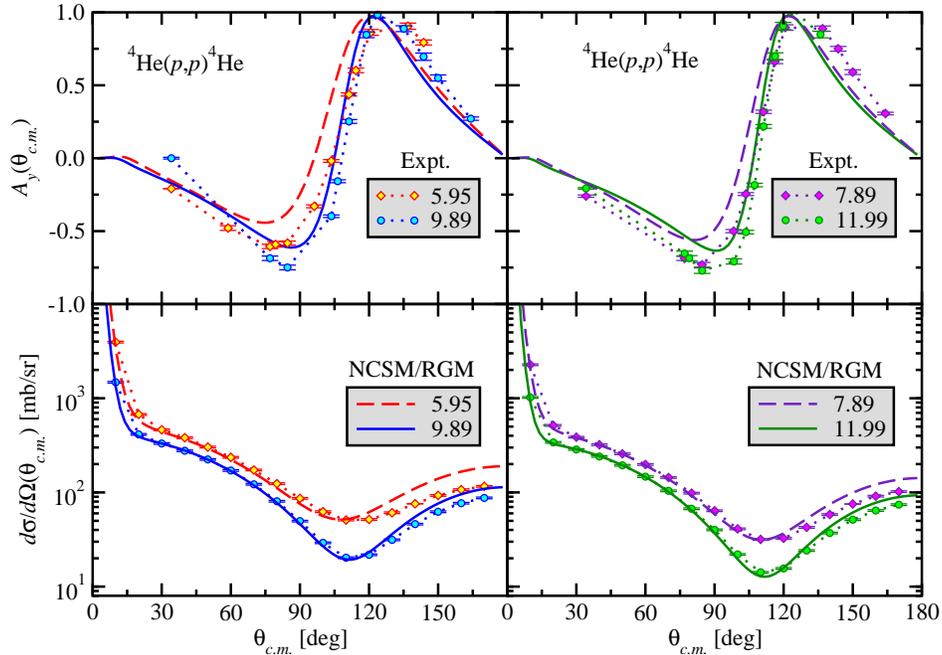}
\caption{(Color online) Calculated $p-^4$He differential cross section (top panels) and  analyzing power (bottom panels) for proton laboratory energies of $E_p = 5.95, 7.89, 9.89$ and 11.99 MeV compared to experimental data from Ref.~\cite{Schwandt}. The NCSM/RGM results include the $^4$He ground state and first $0^+ 0$ excited state and were obtained using the SRG-N$^3$LO $NN$ potential with a cutoff of 2.02 fm$^{-1}$, for an HO frequency $\hbar\Omega=20$ MeV and basis space size $N_{\rm max}=17$.}
\label{fig:p4He_1}
\end{figure*}
\begin{figure*}[t]
\includegraphics*[width=1.0\columnwidth, angle = -90]{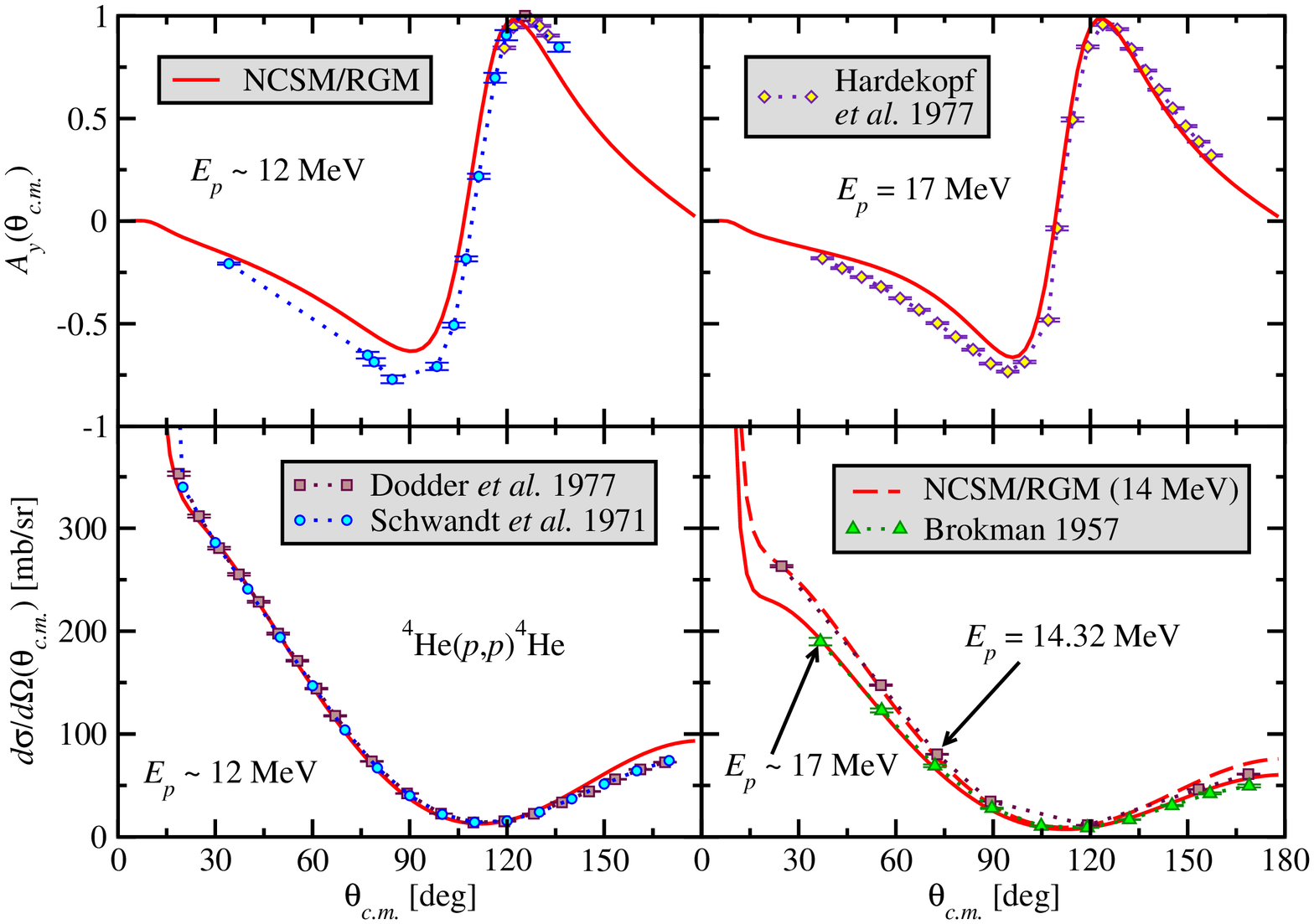}
\caption{(Color online) Calculated $p-^4$He differential cross section (top panels) and  analyzing power (bottom panels) for proton laboratory energies $E_p = 12, 14.32$ and 17 MeV compared to experimental data from Schwandt {\em et al}.~\cite{Schwandt}, Brokman~\cite{Brokman},  Dodder {\em et al}.~\cite{Dodder}, and Hardekopf {\em et al.}~\cite{Hardekopf}. The NCSM/RGM results include the $^4$He ground state and first $0^+ 0$ excited state and were obtained using the SRG-N$^3$LO $NN$ potential with a cutoff of 2.02 fm$^{-1}$, for an HO frequency $\hbar\Omega=20$ MeV and basis space size $N_{\rm max}=17$.}
\label{fig:p4He_2}
\end{figure*}
As our calculated phase shifts agree with the experimental ones reasonably well above the center-of-mass energy of 8 MeV, we expect a similar behavior for cross section and analysing power in that energy range. This is indeed the case as shown in Fig.~\ref{fig:N4He_17MeVxsay}, where the calculated differential cross section and analyzing power are compared to experimental data from Karlsruhe~\cite{Karlsruhe} with polarized neutrons of $E_n=$17 MeV laboratory energy. For the cross section experimental data see also references in~\cite{Karlsruhe}. The cross section is reproduced remarkably well at all angles and the analysing power is in reasonable agreement with the data, particularly at backward angles. The same quality of agreement can be found for all energies far from the low-lying resonances, as shown in the right panel of Fig.~\ref{fig:N4He_17MeVxsay} for the analysing power at $E_n=15$ MeV and  19 MeV.

A better display of the dependence of our calculated cross section and analysing power upon the incident nucleon energy is provided by Fig.~\ref{fig:p4He_1}, where the $p-^4$He results for these observables are compared to the data of Ref.~\cite{Schwandt} at the proton laboratory energies of $E_p = 5.95$, 7.89, 9.89, and 11.99 MeV. As expected from the behavior of the phase shifts described earlier, for energies relatively close to the resonance region we find a rather poor agreement with experiment, particularly noticeable in the analysing power overall and in the cross section at backward angles. However, starting at about 10 MeV, the agreement improves substantially and data are once again reproduced in a quite satisfactory way at higher energies, as shown in Fig.~\ref{fig:p4He_2}, where the NCSM/RGM $p-^4$He results are compared to various experimental data sets~\cite{Schwandt,Brokman,Dodder,Hardekopf} in the energy range $E_p \sim 12-17$ MeV.

\section{Neutron-$^{7}$Li and proton-$^7$Be scattering}
\label{n7Li}

The $^7$Be($p$,$\gamma$)$^8$B capture reaction plays a very important role in nuclear astrophysics as it serves as an input for understanding the solar neutrino flux~\cite{Adelberger}. While the experimental determination of the neutrino flux from $^8$B has an accuracy of about 9\%~\cite{SNO}, the theoretical predictions have uncertainties of the order of 20\%~\cite{CTK03,BP04}. The theoretical neutrino flux depends on the  $^7$Be($p$,$\gamma$)$^8$B S-factor. Significant experimental and theoretical effort has been devoted to studying this reaction. The S-factor extrapolation to astrophysically relevant energies depends among other things on the scattering lengths of the proton scattering on $^7$Be. Experimental determination of these lengths was performed recently~\cite{Be7_scatl} with precision of the order of 30\%. The proton-$^7$Be elastic scattering was also investigated in Ref.~\cite{Rogachev01}. To benchmark the theoretical calculations used for S-factor extrapolations, an investigation of the mirror capture reaction, $^7$Li($n$,$\gamma$)$^8$Li, as well as the $n$+$^7$Li scattering is important. For example, the $n$+$^7$Li scattering lengths are known with a higher accuracy~\cite{Li7_scatl}. 

\begin{figure}[t]
\includegraphics*[width=1.0\columnwidth]{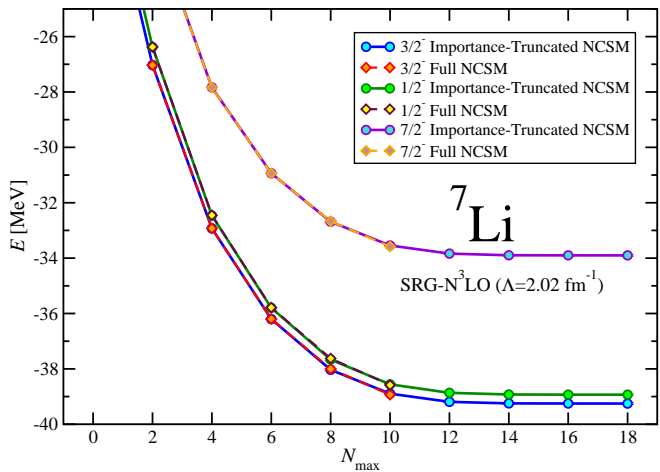}%
\caption{(Color online) $^7$Li ground-state and the $1/2^-$ and $7/2^-$ excited state energy dependence on the model-space size $N_{\rm max}$, obtained within the importance-truncated NCSM (solid lines), using the SRG-N$^3$LO $NN$ potential with a cutoff of 2.02 fm$^{-1}$. The HO frequency $\hbar\Omega=20$ MeV was employed. The full-space NCSM results are shown by dashed lines.}
\label{fig:Li7_ITNCSM}
\end{figure}

The first applications of the NCSM approach to the description of the  $^7$Be($p$,$\gamma$)$^8$B capture reaction~\cite{NBC06} required a phenomenological correction of the asympotic behavior of the overlap functions and, further, the scattering $p$+$^7$Be wave function was calculated from a phenomenological potential model. The present investigation within the {\it ab initio} NCSM/RGM approach paves the way for a complete first principles calculation of this capture reaction. Here, we limit ourselves to scattering calculations and postpone the capture reaction calculations to a forthcoming paper. 

Our current limit on the unrestricted NCSM calculations for $^7$Li and $^7$Be is $N_{\rm max}=10$. To improve the convergence of our scattering calculations, we utilize wave functions obtained within the IT-NCSM. In that scheme, we are able to reach  $N_{\rm max}=18$ model spaces and calculate both ground as well as low-lying excited states. This is demonstrated in Fig.~\ref{fig:Li7_ITNCSM}. With the SRG-N$^3$LO $NN$ potential with $\Lambda=2.02$ fm$^{-1}$ employed in the present study we reach convergence already around $N_{\rm max}=12-14$. Also, as seen in the figure, the aggreement between the unrestricted NCSM and the IT-NCSM is perfect up to the highest accessible unrestricted space, $N_{\rm max}=10$.

\subsection{$n$-$^{7}$Li}

\begin{figure}[t]
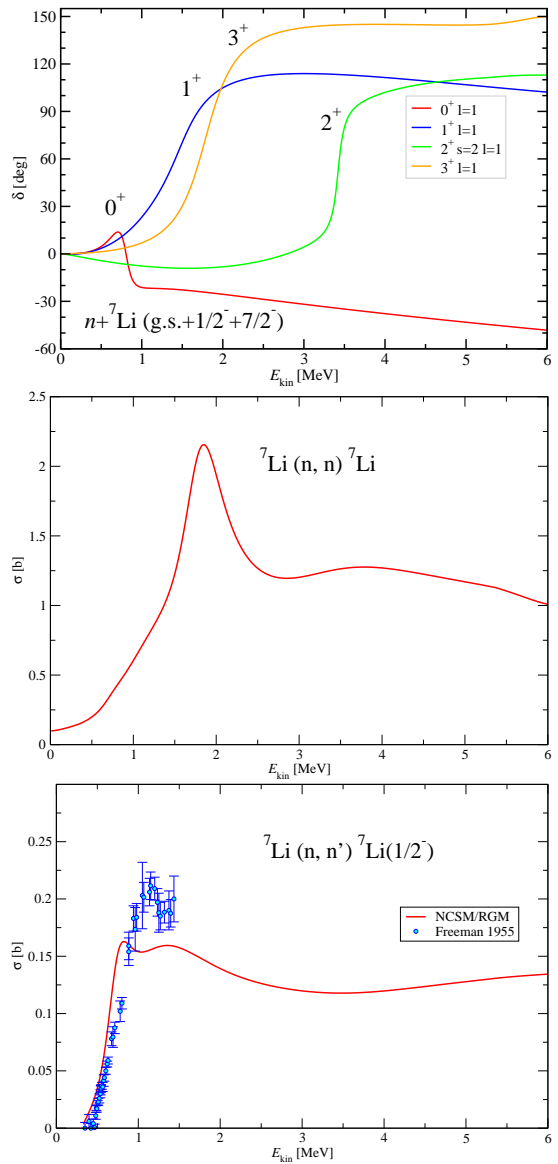

\begin{minipage}{8cm}
\includegraphics*[width=0.9\columnwidth]{phase_shift_nLi7_srg-n3lo0600_20_15_IT_317_Pwaves_fig.eps}%
\end{minipage}
\hfill
\begin{minipage}{8cm}
\includegraphics*[width=0.9\columnwidth]{sigma_reac_nLi7_srg-n3lo0600_20_15_IT_317_2cf_010_ici_gs-gs.eps}%
\end{minipage}
\hfill
\begin{minipage}{8cm}
\includegraphics*[width=0.9\columnwidth]{sigma_reac_nLi7_srg-n3lo0600_20_15_IT_317_2cf_010_ici_gs-1m.eps}%
\end{minipage}
\caption{(Color online) $P$-wave diagonal phase shifts of the $n$-$^7$Li elastic scattering (top panel), elastic $^7$Li($n$,$n$)$^7$Li cross section (middle panel), and inelastic $^7$Li($n$,$n'$)$^7$Li(1/2$^-$) cross section (bottom panel). The NCSM/RGM calculation that included the $^7$Li ground state and the $1/2^-$ and $7/2^-$ excited states were done using the SRG-N$^3$LO $NN$ potential with a cutoff of 2.02 fm$^{-1}$. Wave functions from IT-NCSM calculations in the  $N_{\rm max}=12$ basis and the HO frequency of $\hbar\Omega=20$ MeV were employed. Experimental data are from Ref.~\protect\cite{FLR55}.
}
\label{fig:Li7_IT_317_Pwaves}
\end{figure}
The NCSM/RGM coupled-channel calculations performed for the $A=8$ system include the $^7$Li ($^7$Be) ground state, the first excited $1/2^-$ state as well as the second excited $7/2^-$ state. It is essential to include the $7/2^-$ state in order to reproduce the low-lying $3^+$ resonance in $^8$Li and $^8$B. Using these three states, we are able to reach model spaces up to $N_{\rm max}=12$, which is sufficient concerning the HO basis expansion convergence as can be judged from Fig.~\ref{fig:Li7_ITNCSM}. The coupled channel calculation described above gives two bound states for the $n$-$^7$Li system, a $2^+$ corresponding to the experimentally observed $^8$Li ground state, bound by 2.03 MeV~\cite{TUNL_A8}, and a $1^+$ corresponding to the $^8$Li first excited state at $E_x=0.98$ MeV, bound by 1.05 MeV~\cite{TUNL_A8}. The calculated states are bound by 1.16 MeV and 0.17 MeV, respectively, i.e. less than in experiment. This is in part due to the fact that higher excited states of $^7$Li were omitted. In Fig.~\ref{fig:Li7_IT_317_Pwaves}, we present our results for the diagonal $P$-wave phase shifts of the  $n$+$^7$Li elastic scattering as well as the elastic $^7$Li($n$,$n$)$^7$Li and inelastic $^7$Li($n$,$n'$)$^7$Li(1/2$^-$) cross sections. At low energies, we can identify four resonances two of which can be associated with the experimentally known $^8$Li states: $3^+$ at $E_x=2.255$ MeV and $1^+$ at $E_x=3.21$ MeV~\cite{TUNL_A8}. The other two resonances, $0^+$ and $2^+$ are not present in the $^8$Li evaluation of Ref.~\cite{TUNL_A8}. They do appear in many theoretical calculations including the GFMC~\cite{GFMC}, NCSM~\cite{NBC06} and recoil-corrected continuum shell model (RCCSM)~\cite{Halderson06}. The $0^+$ resonance also appears in the GCM calculations of Ref.~\cite{Desc94}. Contributions of different resonances to the cross sections can be deduced from Fig.~\ref{fig:Li7_IT_317_Pwaves}. The elastic cross section is dominated by the $3^+$ resonance with some contributions from the $2^+$ resonace at higher energy. The inelastic cross section shows a peak just above the  threshold due to the $0^+$ resonance and also a contribution from the $1^+$ resonance. The appearance of a $0^+$ peak just above threshold of the  $^7$Li($n$,$n'$)$^7$Li(1/2$^-$) reaction was also discussed in Ref.~\cite{Halderson06} (see Fig. 10 in that paper). The data of Ref.~\cite{FLR55} seem to rule out a $0^+$ state so close to the threshold. It is known, however, that the position of the $0^+$ state is sensitive to the strength of the spin-orbit interaction~\cite{GFMC,NBC06,Halderson06}. The three-nucleon interaction, that would increase the strenght of the spin-orbit force, was not included in our present calculations. Consequently, our predicted $0^+$ state energy is likely underestimated. We note that no fit to the experimental treshold was done in the present NCSM/RGM calculations. Still, as seen in the bottom panel of Fig.~\ref{fig:Li7_IT_317_Pwaves}, the calculated inelastic cross section is very close to the experimental data just above the threshold.

\subsection{$p$-$^{7}$Be}

\begin{figure}[t]
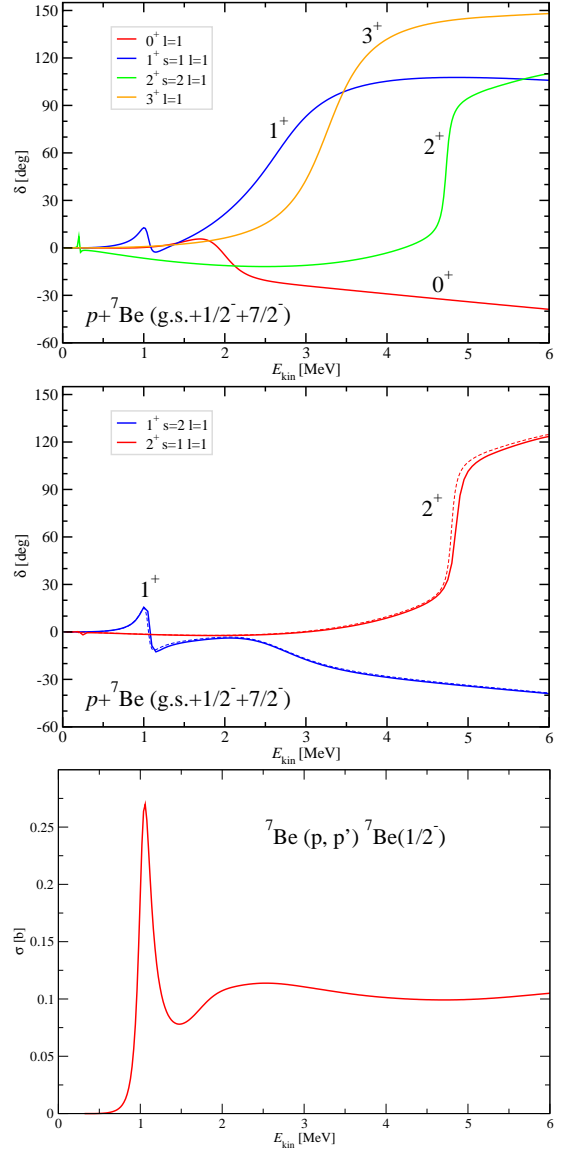

\begin{minipage}{8cm}
\includegraphics*[width=0.9\columnwidth]{phase_shift_pBe7_srg-n3lo2.0205_20_15_317_IT_Pwaves_fig.eps}%
\end{minipage}
\hfill
\begin{minipage}{8cm}
\includegraphics*[width=0.9\columnwidth]{phase_shift_pBe7_srg-n3lo2.0205_20_13_317_Pwaves_full_IT_compare_fig.eps}%
\end{minipage}
\hfill
\begin{minipage}{8cm}
\includegraphics*[width=0.9\columnwidth]{sigma_reac_pBe7_srg-n3lo2.0205_20_15_317_IT_gs-1m.eps}%
\end{minipage}
\caption{(Color online) $P$-wave diagonal phase shifts of the $p$+$^7$Be elastic scattering (top and middle panel) and inelastic $^7$Be($p$,$p'$)$^7$Be(1/2$^-$) cross section (bottom panel). The NCSM/RGM calculation that included the $^7$Be ground state and the $1/2^-$ and $7/2^-$ excited states were done using the SRG-N$^3$LO $NN$ potential with a cutoff of 2.02 fm$^{-1}$. Wave functions from IT-NCSM calculations in the  $N_{\rm max}=12$ basis and the HO frequency of $\hbar\Omega=20$ MeV were employed. In the middle panel, the full-space NCSM (solid lines) and the IT-NCSM (dashed lines) results in the  $N_{\rm max}=10$ basis are compared.}
\label{fig:Be7_317_Pwaves}
\end{figure}
In the mirror system, $p$-$^7$Be, we do not find a bound state in the same type of coupled-channel NCSM/RGM calculation as described above for $n$-$^7$Li. As seen in the top and the middle parts of Fig.~\ref{fig:Be7_317_Pwaves}, the lowest $2^+$ resonance corresponding to the $^8$B ground state lies at about 200 keV above the threshold. In experiment, $^8$B is bound by 137 keV~\cite{TUNL_A8}. Our calculated lowest $1^+$ resonance appears at about 1 MeV. It corresponds to the experimental $^8$B $1^+$ state at $E_x=0.77$ MeV (0.63 MeV above the $p$-$^7$Be threshold). This resonance dominates the inelastic cross section as seen in the bottom part of Fig.~\ref{fig:Be7_317_Pwaves}. The higher lying resonances follow similar patterns as those found in $n$-$^7$Li (Fig.~\ref{fig:Li7_IT_317_Pwaves}). Again, we find $0^+$ and $2^+$ resonances not included in the recent $^8$B evaluation~\cite{TUNL_A8}. We note that experimental efforts are now under way to find these resonances~\cite{Rogachev01,Greife07}. We further note that our calculated $1^+_2$ states in $^8$Li and $^8$B appear at a significantly higher energies than the corresponding $1^+_2$ states obtained within the microscopic cluster model in Ref.~\cite{Csoto}.

The middle panel of Fig.~\ref{fig:Be7_317_Pwaves} demonstrates once again the good accuracy of the importance truncated calculations for a high $N\hbar\Omega$, $N_{\rm max}=10$, model space. The IT calculation reduced the $^7$Be basis from 43.6 million to 11.9 million in the present case.

\begin{figure}[t]
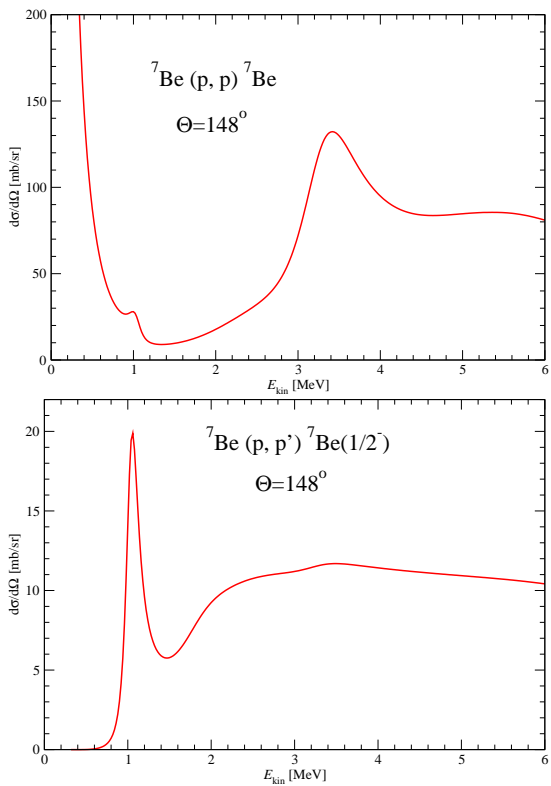

\begin{minipage}{8cm}
  \includegraphics*[width=0.9\columnwidth]{dsigma_dOmega_148_pBe7_srg-n3lo2.0205_20_15_317_IT_gs.eps}
\end{minipage}
\hfill
\begin{minipage}{8cm}
  \includegraphics*[width=0.9\columnwidth]{dsigma_dOmega_148_pBe7_srg-n3lo2.0205_20_15_317_IT_gs-1m.eps}
\end{minipage}
\caption{(Color online) Elastic $^7$Be($p$,$p$)$^7$Be (top panel) and inelastic $^7$Be($p$,$p'$)$^7$Be(1/2$^-$) (bottom panel) differential cross section at $\Theta_{c.m.}=148^0$ calculated within the NCSM/RGM with SRG-N$^3$LO $NN$ potential with $\Lambda=2.02$ fm$^{-1}$.}
\label{fig:p_Be7_148}
\end{figure}
The elastic $p$-$^7$Be scattering was measured at $148^o$ and analyzed by the R-matrix approach~\cite{Rogachev01}. Cross section calculations within the RCCSM at that angle were then published in Ref.~\cite{Halderson04} and also in Ref.~\cite{Halderson06}. Further, elastic and inelastic cross sections at this angle were analyzed within the time-dependent approach to the continuum shell model (TDCSM)~\cite{Volya}. Our elastic and inelastic differential cross section results at $148^o$ are presented in Fig.~\ref{fig:p_Be7_148}. In the elastic cross section, the first $1^+$ state is visible and beyond the minimum of the cross section, we can see the dominant peak due to the $3^+$ state. At higher energies, the $2^+$ state contributes as well. The inelastic cross section at $148^o$ has a similar shape as the reaction cross section shown in Fig.~\ref{fig:Be7_317_Pwaves}. The first $1^+$ state peak dominates at low energy with contributions from the $0^+$ and the second $1^+$ at higher energies. Our findings are in line with the RCCSM results. However, we remind the reader that there is no fitting in our calculations, all results being predictions based on the SRG-N$^3$LO $NN$ potential. Because of this, the positions of our calculated resonances, e.g., $1^+$, $3^+$ do not exactly reproduce experiment. We do not include the experimental data in the figure as they would be shifted compared to the calculated peaks. There are at least two reasons why our predictions do not match the experimental resonances accurately. First, our nuclear Hamiltonian is incomplete, e.g. no three-nucleon interaction is included. Second, we omitted higher resonances of $^7$Li and $^7$Be due to numerical reasons. Most likely, the omitted resonances would produce some shifts in the calculated peaks. Both these points can and will be addressed in the future. Still, our current results contain the bulk of the physics behind the investigated scattering processes. 

\subsection{$S$-wave scattering lengths of $n$-$^{7}$Li and $p$-$^{7}$Be}
\begin{figure}[t]
\includegraphics*[width=1.0\columnwidth]{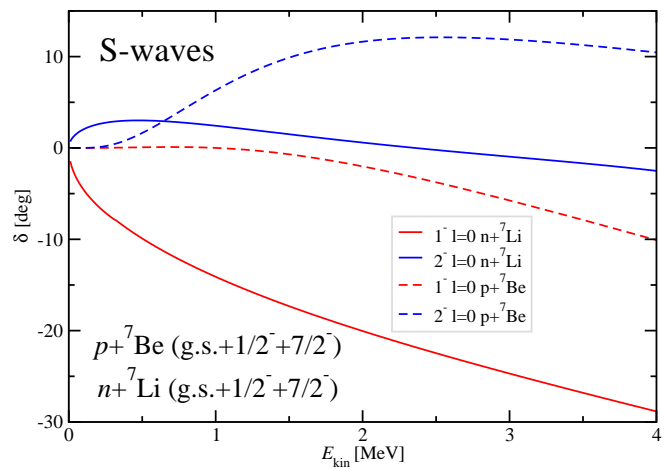}%
\caption{(Color online) $S$-wave phase shifts of the $n$+$^7$Li (solid lines) and the $p$+$^7$Be (dashed lines) elastic scattering. The calculations as described in Figs.~\protect\ref{fig:Li7_IT_317_Pwaves} and~\protect\ref{fig:Be7_317_Pwaves}.}
\label{fig:Be7_Li7_317_Swaves}
\end{figure}
\begin{table}[b]
\begin{ruledtabular}
\begin{tabular}{ccccc}
                      &\multicolumn{2}{c}{$^{7}$Li} &\multicolumn{2}{c}{$^{7}$Be}\\
                      &   Calc.      &  Expt.               &   Calc.        &   Expt.            \\
\hline
$a_{01}$ [fm]   &  +1.23     &  +0.87(7)          &  -1.2         &   25(9)            \\         
$a_{02}$ [fm]   &  -0.61     &  -3.63(5)           &  -10.2       &   -7(3)            \\
\end{tabular}
\end{ruledtabular}
\caption{ The $n$-$^7$Li and the $p$-$^7$Be $S$-wave scattering lengths. Theoretical values correspond to calculations as described in Figs.~\protect\ref{fig:Li7_IT_317_Pwaves} and~\protect\ref{fig:Be7_317_Pwaves}. Experimental values are from Refs.~\protect\cite{Be7_scatl,Li7_scatl}.}\label{tab:S-wave_scatl}
\end{table}
In Fig.~\ref{fig:Be7_Li7_317_Swaves}, we present our calculated $n$-$^7$Li and the $p$-$^7$Be $S$-wave phase shifts. We do not find any evidence for a $2^-$ resonance advocated in Ref.~\cite{Rogachev01} and discussed in Ref.~\cite{Barker00}. The corresponding scattering lengths together with the experimental values are given in Table~\ref{tab:S-wave_scatl}. With the exception of the $p$-$^7$Be $a_{01}$, which has a large experimental uncertainty, our calculated scattering lengths do agree with experimental data as to their signs, there are however differences in the absolute values. Again, as dicussed above, the results presented here serve only as a first step towards the {\it ab initio} investigation of the $n$-$^7$Li and the $p$-$^7$Be reactions. Prospects for a realistic calculation of the $^7$Be($p$,$\gamma$)$^8$B capture are excellent. Here we found the $^8$B unbound by only 200 keV. It is quite possible that $^8$B will become bound (with the $NN$ potential employed here: SRG-N$^3$LO with $\Lambda=2.02$ fm$^{-1}$) by including more excited states of $^7$Be in the coupled-channel NCSM/RGM calculations. The effect of higher excited states of $^7$Be can be, in fact, most efficiently included by coupling the presently used NCSM/RGM basis with the $^8$B NCSM eigenstates as outlined in Ref.~\cite{NCSM_review}. Even if the $^8$B would not be bound or, most likely, the theshold energy will not agree with the experiment, we have the possibilty to explore a variation of the SRG $NN$ potential evolution parameter $\Lambda$ and tune this parameter to fit the experimental threshold. We note that for any $\Lambda$ the SRG-evolved $NN$ potential will describe all two-nucleon properties as accurately as the original starting $NN$ potential, here the chiral N$^3$LO potential of Ref.~\cite{N3LO}. It should be noted that by adding the three-nucleon interaction, omitted in the present calculations due to computational reasons, the need for a fine-tuning should be significantly reduced, i.e. the results should become $\Lambda$ independent.

\section{Nucleon-$^{12}$C scattering}
\label{n12C}

\begin{figure}[t]
\includegraphics*[width=1.0\columnwidth]{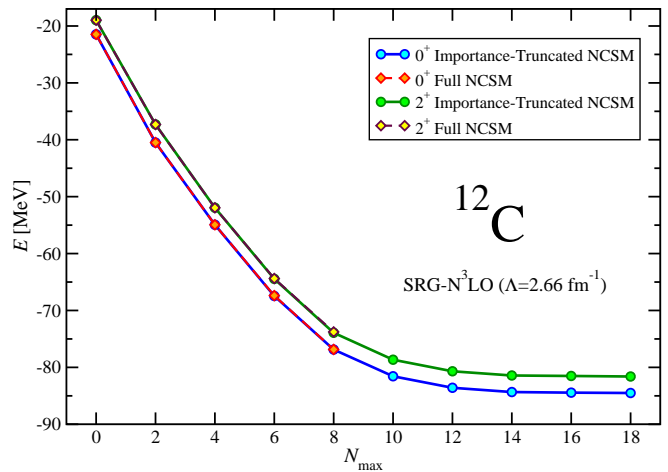}%
\caption{(Color online) Ground-state and the first excited $2^+$ state energy dependence on the model-space size $N_{\rm max}$ for $^{12}$C, obtained within the importance-truncated NCSM, using the SRG-N$^3$LO $NN$ potential with a cutoff of 2.66 fm$^{-1}$. The HO frequency $\hbar\Omega=24$ MeV was employed. The calculation is variational. No NCSM effective interaction was used. The full NCSM results were obtained with the code Antoine~\cite{Antoine}.}
\label{fig:C12_ITNCSM}
\end{figure}

For nucleon scattering calculations on $^{12}$C or heavier targets within the NCSM/RGM, the use of the importance truncation becomes essential. For $^{12}$C, the full-space NCSM calculations are currently limited to $N_{\rm max}=8$ (although successful runs were already performed for $N_{\rm max}=10$ on the biggest supercomputers with the latest version of the code MFD~\cite{MFD}). This is insufficient for reaching or approaching convergence of the $^{12}$C NCSM calculations as seen from Fig.~\ref{fig:C12_ITNCSM} and even more so of the NCSM/RGM scattering calculations.  The importance-truncated calculations, on the other hand, are feasible up to $N_{\rm max}=18$, where convergence is reached for both the ground state as well excited states. Our $^{12}$C calculations are performed with the SRG-N$^3$LO $NN$ potential with the evolution parameter $\Lambda=2.66$ fm$^{-1}$, a higher value (i.e. shorter evolution, less soft) than that used for the lighter nuclei. The use of a small $\Lambda$ results in large overbinding of heavier nuclei and a significant underestimation of their radii. As seen in Fig.~\ref{fig:C12_ITNCSM}, our converged $^{12}$C binding energy is about 84.5(8) MeV, smaller than the experimental value of 92 MeV and, further, the agreement of the full-space and importance-truncated results is perfect all the way up to $N_{\rm max}=8$.  

\subsection{$p$-$^{12}$C}

\begin{figure}[t]
\includegraphics*[width=1.0\columnwidth]{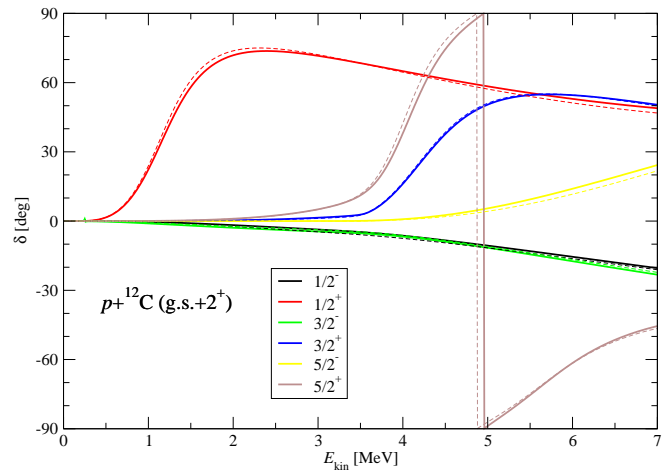}
\caption{(Color online) The $p$-$^{12}$C eigenphase shifts calculated within the NCSM/RGM using the SRG-N$^3$LO $NN$ potential with a cutoff of 2.66 fm$^{-1}$ and the  HO frequency $\hbar\Omega=24$ MeV. Full lines (dotted lines) corespond to results obtained in the $N_{\rm max}=16$ ($N_{\rm max}=14$) model space. The ground state and the first excited $2^+$ state of $^{12}$C was included. The $^{12}$C wave functions were obtained within the IT NCSM.}
\label{fig:pC12}
\end{figure}

Our low-energy $p$-$^{12}$C phase shift results are shown in Fig.~\ref{fig:pC12}. The comparison of the $N_{\rm max}=16$ and $N_{\rm max}=14$ results demonstrates good convergence with respect to the HO basis expansion.  The $^{12}$C ground state and the first $2^+$ state were included in the coupled-channels NCSM/RGM equations. We note that we also performed a phase shift comparison of the full-space and the importance-truncated calculations up to $N_{\rm max}=6$ and found a similarly perfect agreement as presented in Fig.~\ref{fig:nHe4_phase_full_IT} for $n$-$^4$He.
In the present $p$-$^{12}$C calculations, we found a single bound state, $1/2^-$ at -2.98 MeV, corresponding to the $^{13}$N ground state, bound experimentally by 1.94 MeV~\cite{AS91}. The lowest resonance in our calculation is $3/2^-$, barely visible at 0.25 MeV above threshold. In experiment, this resonance is at 1.56 MeV. Our calculated $1/2^+$ resonance appears at about 1.5 MeV above treshold (in experiment at 0.42 MeV above threshold)  and the $5/2^+$ resonance at about 4.9 MeV (in experiment at 2.61 MeV).  

\subsection{$n$-$^{12}$C}

In the mirror system, $n$-$^{12}$C, our NCSM/RGM calculations produce three bound states: $1/2^-$ at -5.34 MeV corresponding to the $^{13}$C ground state experimentally bound by 4.95 MeV with respect to the $n$-$^{12}$C threshold, $3/2^-$, bound by 2.23 MeV (experimentally bound by 1.26 MeV), and $1/2^+$ bound by just 0.03 MeV (experimentally bound by 1.86 MeV). In experiment, there is also a $5/2^+$ state bound by 1.09 MeV. Our present NCSM/RGM calculations including the lowest $0^+$ and and the lowest $2^+$ $^{12}$C states do not produce any bound $5/2^+$ state. 

\begin{figure}[t]
\includegraphics*[width=1.0\columnwidth]{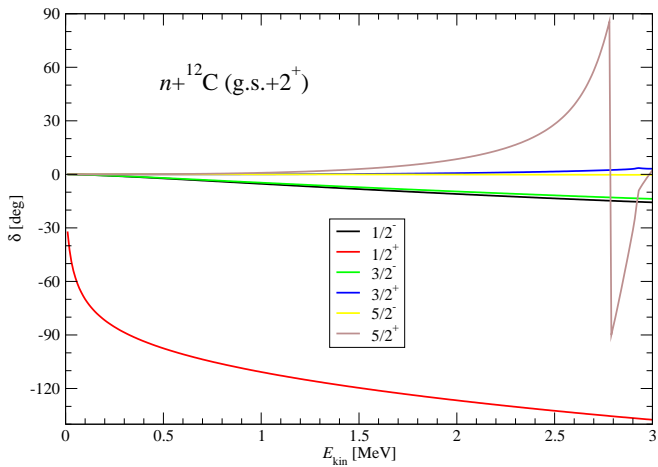}
\caption{(Color online) The $n$-$^{12}$C phase shifts calculated within the NCSM/RGM using the SRG-N$^3$LO $NN$ potential with a cutoff of 2.66 fm$^{-1}$. The HO frequency $\hbar\Omega=24$ MeV and the model-spaces size of $N_{\rm max}=16$ were used. The ground state and the first excited $2^+$ state of $^{12}$C was included. The $^{12}$C wave functions were obtained within the IT NCSM.}
\label{fig:nC12}
\end{figure}
Our low-energy $n$-$^{12}$C diagonal phase shifts are shown in Fig.~\ref{fig:nC12}. The $5/2^+$ resonance is found at 2.8 MeV (experimenally at 1.92 MeV with respect to the $n$-$^{12}$C threshold). The steep drop of the $1/2^+$ phase shift is due to the presence of the very weakly bound $1/2^+$ state. We note that similarly as in the case of $^{11}$Be, discussed in Ref.~\cite{NCSMRGM}, we observe a significant decrease of the $1/2^+$ state energy in the $n$-$^{12}$C NCSM/RGM calculation when compared to the standard NCSM calculation for $^{13}$C. We were able to make these comparisons in model spaces up to $N_{\rm max}=6$ where we found this drop to be about 3 MeV. 

\begin{figure}[t]
\includegraphics*[width=1.0\columnwidth]{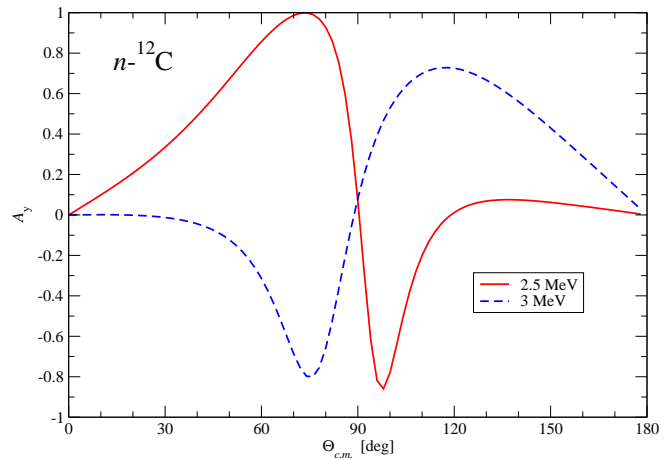}
\caption{(Color online) The analysing power for $n$-$^{12}$C elastic scattering below and above the calculated $5/2^+$ resonance. Energies are in the center of mass. The calculation as described in Fig.~\protect\ref{fig:nC12}.} 
\label{fig:nC12_Ay}
\end{figure}
Analysing powers were measured for proton and neutron scattering on $^{12}$C~\cite{Tra67,Hsu66,Ro05} and scattering experiments on polarized proton target are under way~\cite{GUpriv}. In Fig.~\ref{fig:nC12_Ay}, we present our calculated analyzing power below and above the energy of the $5/2^+$ resonance. We note that our calculated $5/2^+$ resonance appears at 2.8 MeV in the center of mass (experimentally at 1.92 MeV). Below the resonance, the analyzing power is positive at $\Theta_{\rm CM}<90^o$ and negative at $\Theta_{\rm CM}>90^o$. At energies above the resonance, the analyzing power reverses its sign. Similar observations were made in calculations performed within the multichannel algebraic scattering (MCAS) theory~\cite{n_C12_p_C12_MCAS,n_C12_MCAS}. See in particular Fig. 5 of Ref.~\cite{n_C12_MCAS}.

Our calculated $^{13}$N and $^{13}$C bound-state levels and resonances are more spread than the experimental ones. This is a consequence of an underestimation of the $^{12}$C radius found to be 2.05 fm with the SRG-N$^3$LO $NN$ potential. To remedy this, one would have to calculate three-nucleon interaction terms induced due to the SRG evolution. This can be done as described in Ref.~\cite{JNF09}. However, we still need to further develop the NCSM/RGM formalism in order to handle three-nucleon interactions in the scattering calculations.

\section{Nucleon-$^{16}$O scattering}
\label{n16O}

The calculation of nucleon scattering on $^{16}$O is the most challenging among the systems we investigate in this paper. The $\alpha$ clustering plays an important role in the structure of $^{16}$O, in particular for the first excited $0^+$ state that is known to be almost impossible to reproduce in NCSM or coupled-cluster calculations. Our present calculations do not include the $\alpha$ clustering yet. 

\begin{figure}[t]
\includegraphics*[width=1.0\columnwidth]{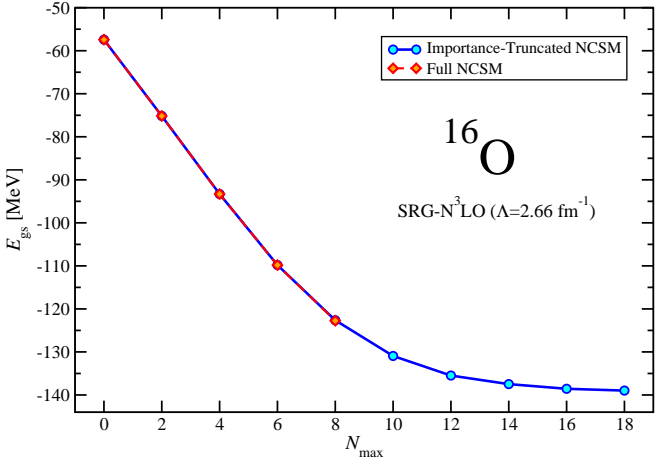}%
\caption{(Color online) Ground-state energy dependence on the model-space size $N_{\rm max}$ for $^{16}$O, obtained within the importance-truncated NCSM, using the SRG-N$^3$LO $NN$ potential with a cutoff of 2.66 fm$^{-1}$. The HO frequency $\hbar\Omega=24$ MeV was employed. The calculation is variational. No NCSM effective interaction was used. The full NCSM results were obtained with the code Antoine~\cite{Antoine}.}
\label{fig:O16_ITNCSM}
\end{figure}

As in the case of $^{12}$C, we rely on the importance-truncated NCSM calculations for obtaining the $^{16}$O wave functions as the full-$N_{\rm max}$ NCSM calculations are possible only up to $N_{\rm max}=8$. In Fig.~\ref{fig:O16_ITNCSM}, we show the ground-state convergence within the IT-NCSM and a comparison to the full-space results. Again, up to the largest accessible model space, the agreement between the importance-truncated and the full-space calculations is perfect.

\subsection{$n$-$^{16}$O}

\begin{figure}[t]
\includegraphics*[width=1.0\columnwidth]{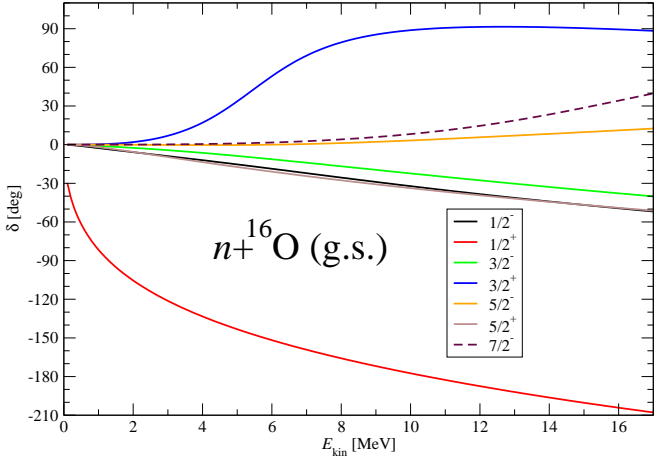}
\caption{(Color online) The $n-^{16}$O phase shifts calculated within the NCSM/RGM using the SRG-N$^3$LO $NN$ potential with a cutoff of 2.66 fm$^{-1}$ and the HO frequency $\hbar\Omega=24$ MeV in the $N_{\rm max}=18$ model space. The ground state and of $^{16}$O was included. The $^{16}$O wave functions were obtained within the IT NCSM.}
\label{fig:nO16gs_18}
\end{figure}
\begin{figure}[t]
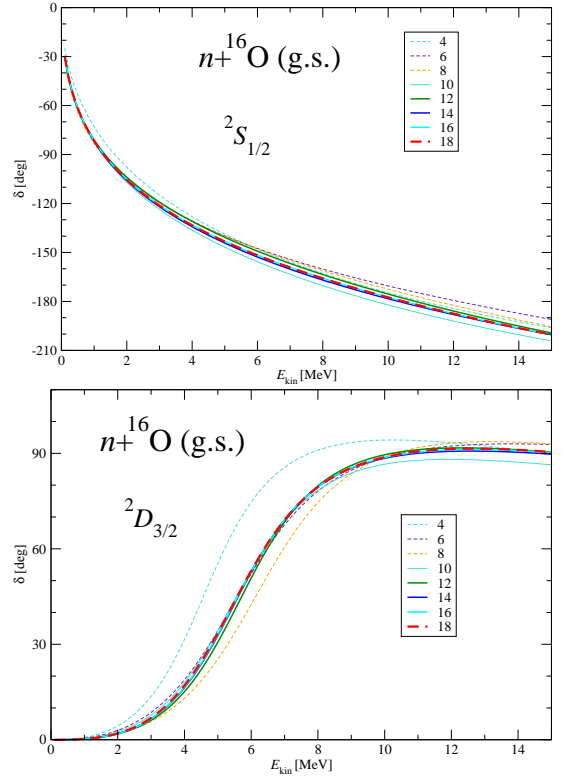

\begin{minipage}{8cm}
  \includegraphics*[width=0.9\columnwidth]{phase_shift_nO16_srg-n3lo0200_24_7_21_1p.eps}
\end{minipage}
\hfill
\begin{minipage}{8cm}
  \includegraphics*[width=0.9\columnwidth]{phase_shift_nO16_srg-n3lo0200_24_7_21_3p.eps}
\end{minipage}
\caption{(Color online) Basis size dependence of the $n-^{16}$O phase shifts calculated within the NCSM/RGM using the SRG-N$^3$LO $NN$ potential with a cutoff of 2.66 fm$^{-1}$. The HO frequency of $\hbar\Omega=24$ MeV was used. The $J^\pi=1/2^+ (3/2^+)$ channel is shown in the top (bottom) panel. Model space sizes up to  $N_{\rm max}=18$ were considered. The ground state and of $^{16}$O was included. The $^{16}$O wave functions were obtained within the IT NCSM.}
  \label{fig:nO16gs_conv}
\end{figure}
It is straightforward to converge nucleon-$^{16}$O scattering calculations within the NCSM/RGM using the HO expansion up to $N_{\rm max}=18$. Our calculated $n$-$^{16}$O phase shifts are shown in Fig.~\ref{fig:nO16gs_18} and the HO-basis expansion convegence is checked for the $S$- and the $D$-wave in Fig.~\ref{fig:nO16gs_conv}. These calculations included the $^{16}$O ground state only. We find two bound states, $1/2^+$ at -0.88 MeV and $5/2^+$ at -0.41 MeV with respect to the $n$-$^{16}$O threshold. In experiment, the $^{17}$O ground state is $5/2^+$, bound by 4.14 MeV, and the $1/2^+$ state is the first excited state bound by 3.27 MeV. There are also two additional bound states: $1/2^-$ and $3/2^-$. Those are unbound in our calculations. 

\begin{figure}[t]
\includegraphics*[width=1.0\columnwidth]{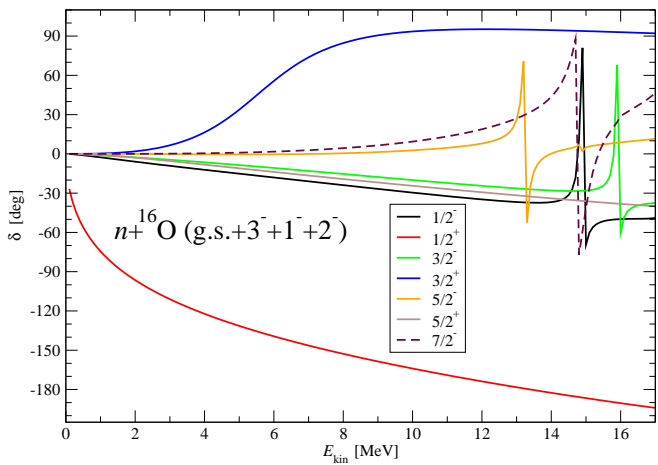}
\caption{(Color online) The $n$-$^{16}$O phase shifts calculated within the NCSM/RGM using the SRG-N$^3$LO $NN$ potential with a cutoff of 2.66 fm$^{-1}$ and the HO frequency $\hbar\Omega=24$ MeV in the $N_{\rm max}=13$ model space. The ground state and and the lowest $3^-$, $1^-$ and $2^-$ excited states of $^{16}$O were included. The $^{16}$O wave functions were obtained within the IT NCSM.}
\label{fig:nO16gs312_15}
\end{figure}
Clearly, it is insufficient to consider only the ground state of $^{16}$O in the coupled-channel NCSM/RGM scattering calculations. We, therefore, include in addition the three lowest $^{16}$O negative parity states: $3^-$, $1^-$, and $2^-$. Due to computational limitations, in this case we used HO basis expansion up to $N_{\rm max}$=13. Comparing Fig.~\ref{fig:nO16gs312_15} to Fig.~\ref{fig:nO16gs_18}, the $1p-1h$ negative-parity excited states of $^{16}$O generate negative-parity resonances in $^{17}$O. These resonances do appear, however, at much higher energy than in experiment. The reason for this is the fact that our calculated $^{16}$O $1p-1h$ states have too large excitation energy. In particular, our calculated $3^-$ excited state has an excitation energy of 15.99 MeV while experimentally it lies at just 6.13 MeV. One reason for the discrepancy is the softness of the SRG-N$^3$LO $NN$ potential we use that results in an overall overbinding of the $^{16}$O ground state and in an underestimation of its radius. Another aspect is the challenging problem of the IT-NCSM extrapolations of the independent positive and negative-parity state calculations. The uncertainties of the relative excitation energies are higher than in same-parity calculations. On the positive side our calculation with the negative-parity states, even though with overestimated excitation energies, results in the proper ordering of the $^{17}$O bound states. The ground state is $5/2^+$ at -1.32 MeV and the $1/2^+$ state gains binding as well, appearing at -1.03 MeV.

\subsection{$p$-$^{16}$O}

We also investigated the $p$-$^{16}$O scattering and $^{17}$F states. When the NCSM/RGM calculations are restricted to the channels involving only the $^{16}$O ground state, we find a $1/2^+$ resonance at 1.0 MeV and a $5/2^+$ resonance at 2.2 MeV. These resonances correspond to the $^{17}$F $1/2^+$ first excited state, bound by 0.105 MeV, and the $^{17}$F $5/2^+$ ground state bound by 0.6 MeV with respect to the $p$+$^{16}$O threshold. By coupling channels involving the $1p-1h$ $^{16}$O $3^-$, $1^-$ and $2^-$ excited states, the calculated $1/2^+$ and $5/2^+$ states are still unbound resonances but their energy moves significantly closer to the threshold: the $1/2^+$ appears at +0.7 MeV and the $5/2^+$ at +1.2 MeV.  

The $^{17}$F low-lying states were recently investigated within the coupled-cluster approach with the Gamow-Hartree-Fock basis~\cite{HPH10}. In those calculations with the N$^3$LO $NN$ potential, the $1/2^+$ state is weakly bound while the $5/2^+$ state remains unbound by about 0.1 MeV. Using the SRG evolved interaction, the $5/2^+$ state became bound with the decrease of the cutoff $\Lambda$. We note that our calculated $^{16}$O ground state energy, -139.0(8) MeV (Fig.~\ref{fig:O16_ITNCSM}) obtained with the SRG-N$^3$LO $NN$ potential with $\Lambda=2.66$ fm$^{-1}$, compares well with the CCSD coupled-cluster $^{16}$O calculations: -137.6 MeV with the SRG-N$^3$LO $NN$ potential with $\Lambda=2.8$ fm$^{-1}$~\cite{Pap10}. The differences in the positions of the $1/2^+$ and the $5/2^+$ are due to deficiencies in our description of the negative parity  $1p-1h$ states, which could be related to the two-body Hamiltonian used here as well as the uncertainties of the threshold extrapolations for the excitation energies. The inclusion of additional $^{16}$O excited states would increase the absolute energy of our calculated $^{17}$F states. The most efficient way to do this is by coupling the presently used NCSM/RGM basis with the $^{17}$F NCSM eigenstates in as outlined in Ref.~\cite{NCSM_review}.

\section{Conclusions}
\label{conclusions}

By combining the  importance truncation scheme for the cluster eigenstate basis with the {\it ab initio} NCSM/RGM approach, we were able to perform many-body calculations for nucleon scattering on nuclei with mass number as high as $A=16$. With the soft SRG-evolved chiral $NN$ potentials, convergence of the calculations with respect to the HO basis expansion of the target eigenstates and the localized parts of the NCSM/RGM integration kernels can be reached using $N_{\rm max}=12-16$. 

We first benchmarked the IT-NCSM results with the full-space NCSM results for the $A=5$ system. Our neutron-$^4$He and proton-$^4$He calculations compare well with an R-matrix analysis of the data in particular at energies above 8 MeV, and describe well measured cross sections and analysing powers for those energies.

Our calculations of  $n$-$^7$Li and $p$-$^7$Be scattering predict low-lying $0^+$ and $2^+$ resonances in $^8$Li and $^8$B that have not been experimentally clearly identified yet. We found that the prospects of a realistic {\it ab initio} calculation of the $^7$Be($p$,$\gamma$)$^8$B capture within our approach are very good. In the present calculations we found the $^8$B unbound by only 200 keV. It is quite possible that $^8$B will become bound (with the $NN$ potential employed here: SRG-N$^3$LO with $\Lambda=2.02$ fm$^{-1}$) by including more excited states of $^7$Be in the coupled-channel NCSM/RGM calculations. Even if the $^8$B will still not be bound or, most likely, the threshold energy will not agree with the experiment, we have the possibility to explore a variation of the SRG $NN$ potential evolution parameter $\Lambda$ and tune this parameter to fit the experimental threshold.

The use of the importance-trunacted basis becomes essential in calculations with $^{12}$C or $^{16}$O targets as the full-space NCSM calculations are limited to $N_{\rm max}=8$. Our $n$-$^{12}$C and $p$-$^{12}$C investigations included $^{12}$C ground and the first excited $2^+$ states. We found a single bound state, $1/2^+$ in $^{13}$N as in experiment. In $^{13}$C, we found three bound states with the $5/2^+$ state still unbound contrary to experiment. Our calculated spectrum of $A=13$ states is more spread than in experiment due to the underestimation of the $^{12}$C radius, a consequence of the softness of the SRG-evolved $NN$ interaction. 

The description of nucleon scattering on $^{16}$O within our formalism was the most challenging. The $\alpha$ clustering that plays an important role in the structure of $^{16}$O is not yet included in our present calculations. Further, the $1p-1h$ $^{16}$O excited states are more difficult to treat in the IT-NCSM approach, as the extrapolations of excitation energies are done from the independent ground state and the negative-parity state calculations. We found a strong impact of the $1p-1h$ $^{16}$O states on the positions of the lowest $A=17$ states. For example, correct ordering of the $5/2^+$ and the $1/2^+$ states in $^{17}$O was obtained only when the $1p-1h$ states were included.

Overall, we find that the inclusion of additional excited states of the target nuclei would be beneficial in all studied systems and more significant with the increase of $A$. Coupled-channel NCSM/RGM calculations with many excited states of the traget are computationally challenging. The most efficient way of including the effects of such states is by coupling the presently used NCSM/RGM basis, consisting of just a few lowest excited states, with the NCSM eigenstates of the composite system as outlined in Ref.~\cite{NCSM_review}. Work on this coupling is under way.

The use of the SRG-evolved $NN$ interaction facilitates convergence of the NCSM/RGM calculations with respect to the HO basis expansion. On the other hand, due to the softness of these interactions, radii of heavier nuclei become underestimated. To remedy this, one would have to calculate three-nucleon interaction terms induced due to the SRG evolution. This can be done as described in Ref.~\cite{JNF09}. It is essential to further develop the NCSM/RGM formalism in order to handle three-nucleon interactions, both genuine and the SRG-evolution induced, in the scattering calculations.

In the present paper, we limited ourselves to single-nucleon projectile scattering. Extensions of the NCSM/RGM formalism to include deuteron, $^3$H and $^3$He projectiles are under way.

\acknowledgments
Numerical calculations have been performed at the LLNL LC facilities and at the NIC, J\"{u}lich. Prepared in part by LLNL under Contract DE-AC52-07NA27344.
Support from the U.\ S.\ DOE/SC/NP (Work Proposal No.\ SCW0498), LLNL LDRD grant PLS-09-ERD-020, and from the U.\ S.\ Department of Energy Grant DE-FC02-07ER41457 is acknowledged. This work is supported in part by the Deutsche Forschungsgemeinschaft through contract SFB 634 and by the Helmholtz International Center for FAIR within the framework of the LOEWE program launched by the State of Hesse. \\


\begin{thebibliography}{10}

\bibitem{benchmark} H. Kamada {\it et al.} Phys. Rev. C {\bf 64}, 044001 (2001).

\bibitem{Nogga00} A. Nogga, H. Kamada, and W. Gl\"ockle, Phys. Rev. Lett. {\bf 85}, 944 (2000). 

\bibitem {GFMC} R. B. Wiringa, S. C. Pieper, J. Carlson, V. R. Pandharipande,
               Phys. Rev. C {\bf 62}, 014001 (2000);
               S. C. Pieper and R. B. Wiringa, Ann. Rev. Nucl. Part. Sci.
               {\bf 51}, 53 (2001);
               S. C. Pieper, K. Varga and R. B. Wiringa, 
               Phys. Rev. C {\bf 66}, 044310 (2002).

\bibitem{NO03} P.\ Navr\'atil and W.\ E.\ Ormand, Phys. Rev. C {\bf 68}, 034305 (2003).

\bibitem{Witala01} H. Witala, W. Gl\"ockle, J. Golak, A. Nogga, H. Kamada, R. Skibinski, 
     and J. Kuros-Zolnierczuk, Phys. Rev. C {\bf 63}, 024007 (2001).

\bibitem{Lazauskas05} R. Lazauskas and J. Carbonell, Phys. Rev. C {\bf 70}, 044002 (2004).

\bibitem{Pisa} A. Kievsky, S. Rosati, M. Viviani, L. E. Marcucci and L. Girlanda, 
               J. Phys. G {\bf 35}, 063101 (2008).

\bibitem{Deltuva} A.\ Deltuva and A.\ C.\ Fonseca, 
                      Phys.\ Rev.\ C {\bf 75}, 014005 (2007);  
                      Phys.\ Rev.\ Lett. {\bf 98}, 162502 (2007).

\bibitem{GFMC_nHe4} K.\ M.\ Nollett, S. C. Pieper, R. B. Wiringa, 
                    J. Carlson and G. M. Hale, 
                    Phys.\ Rev.\ Lett.\ {\bf 99}, 022502 (2007). 

\bibitem{Ha07} G. Hagen, D. J. Dean, M. Hjorth-Jensen and T. Papenbrock, Phys. Lett. B {\bf 656}, 169 (2007).

\bibitem{NCSMC12} P.\ Navr\'atil, J.\ P.\ Vary, and B.\ R.\ Barrett,
                   Phys.\ Rev.\ Lett. {\bf 84}, 5728 (2000);
                   Phys.\ Rev.\ C {\bf 62}, 054311 (2000).

\bibitem{RGM} K.\ Wildermuth and Y.\ C.\ Tang, {\it A unified theory of the nucleus},
              (Vieweg, Braunschweig, 1977). 

\bibitem{RGM1} Y.\ C.\ Tang, M. LeMere and D. R. Thompson,
              Phys.\ Rep.\ {\bf 47}, 167 (1978).
             
\bibitem{RGM2} T. Fliessbach and H. Walliser, Nucl. Phys. {\bf A377}, 84 (1982).

\bibitem{RGM3} K.\ Langanke and H.\ Friedrich, {\it Advances in Nuclear Physics}, 
              edited by J. W. Negele and E. Vogt (Plenum, New York, 1986).

\bibitem{Lovas98} R. G. Lovas, R. J. Liotta, A. Insolia, K. Varga and D. S. Delion,
              Phys. Rep. {\bf 294}, 265 (1998).

\bibitem{Hofmann08} H. M. Hofmann and G. M. Hale, Phys. Rev. C {\bf 77}, 044002 (2008).

\bibitem{NCSMRGM} S.\ Quaglioni and P.\ Navr{\'a}til,  Phys. Rev. Lett. {\bf 101}, 092501 (2008). 

\bibitem{NCSMRGM_PRC} S.\ Quaglioni and P.\ Navr{\'a}til,  Phys. Rev. C  {\bf 79}, 044606 (2009).
                     
\bibitem{3bbound1} P.\ Descouvemont, C.\ Daniel, and D.\ Baye, Phys.\ Rev.\  C {\bf 67}, 044309 (2003).

\bibitem{3bcont1} P.\ Descouvemont, E.\ Tursunov, and D.\ Baye, Nucl.\ Phys.\ {\bf A765}, 370 (2006).

\bibitem{3bcnfr} M.\ Theeten, D.\ Baye, and P.\ Descouvemont, Phys.\ Rev.\ C {\bf 74}, 044304 (2006).

\bibitem{3bbound2} M.\ Theeten, H.\ Matsumura, M.\ Orabi, D.\ Baye, P.\ Descouvemont, Y.\ Fujiwara, and Y.\ Suzuki, Phys.\ Rev.\ C {\bf 76}, 054003 (2007).

\bibitem{3bcont2} D.\ Baye, P.\ Capel, P.\ Descouvemont, and Y.\ Suzuki, Phys. Rev. C {\bf 79}, 024607 (2009).

\bibitem{IT-NCSM} R. Roth and P. Navr\'atil, Phys. Rev. Lett. {\bf 99}, 092501 (2007).

\bibitem{Roth09} R. Roth, Phys. Rev. C {\bf 79}, 064324 (2009).

\bibitem{SRG} S. K. Bogner, R. J. Furnstahl and R. J. Perry, Phys. Rev. C {\bf 75}, 061001 (2007).

\bibitem{Roth_SRG} R. Roth, S. Reinhardt and H. Hergert, Phys. Rev. C {\bf 77}, 064003 (2008). 

\bibitem{Roth_PPNP} R. Roth, T. Neff, H. Feldmeier, Prog. Part. Nucl. Phys. \textbf{65}, 50 (2010).

\bibitem{N3LO} D.\ R.\ Entem and R.\ Machleidt, Phys.\ Rev.\ C {\bf 68}, 041001(R) (2003).
 
\bibitem{RoGo09} R. Roth, J. R. Gour, and P. Piecuch, Phys. Rev. C \textbf{79}, 054325 (2009).

\bibitem{RoGo09b} R. Roth, J. R. Gour, and P. Piecuch, Phys. Lett. \textbf{B} 682, 27 (2009).

\bibitem{BoKu03} S.\ K.\ Bogner, T.\ T.\ S.\ Kuo, and A.\ Schwenk, 
                                 Phys.\ Rept.\ \textbf{386}, 1 (2003); 
                                 G.\ Hagen, private communication.

\bibitem{CD-Bonn2000} R.\ Machleidt, Phys.\ Rev.\ C {\bf 63}, 024001 (2001). 

\bibitem{HalePriv} G.\ M.\ Hale, private communication.

\bibitem{Karlsruhe} H. Krupp, J. C. Hiebert, H. O. Klages, P. Doll, J. Hansmeyer, P. Klischke, J. Wilczynski, 
                 and H. Zankel, Phys. Rev. C {\bf 30}, 1810 (1984).

\bibitem{Schwandt} P. Schwandt, T. B. Clegg, and W. Haeberli, Nucl. Phys. A {\bf 163}, 432 (1971).

\bibitem{Brokman} K. W. Brokman, Phys. Rev. {\bf 108}, 1000 (1957).                 

\bibitem{Dodder} D. C. Dodder, G. M. Hale, N. Jarmie, J. H. Jett, P. W. Keaton, Jr., R. A. Nisley, and K. Witte, Phys. Rev. C {\bf 15}, 518 (1977).

\bibitem{Hardekopf} R. A. Hardekopf and G. G. Holsen, Phys. Rev. C {\bf 15}, 514 (1977).

\bibitem{Adelberger} E. Adelberger {\it et al.}, rev. Mod. Phys. {\bf 70}, 1265 (1998).

\bibitem{SNO} SNO Collaboration, S. N. Ahmed {\it et al.}, Phys. Rev. Lett. {\bf 92},
              181301 (2004).

\bibitem{CTK03} S. Couvidat, S. Turck-Chi\`eze, and A. G. Kosovichev, Astrophys. J. {\bf 599},
                1434 (2003).

\bibitem{BP04} J. N. Bahcall and M. H. Pinsonneault, Phys. Rev. Lett. {\bf 92}, 121301 (2004).

\bibitem{Be7_scatl} C. Angulo {\it et al.}, Nucl. Phys. A {\bf 716}, 211 (2003).

\bibitem{Rogachev01} G. V. Rogachev {\it et al.}, Phys. Rev. C {\bf 64}, 061601(R) (2001).

\bibitem{Li7_scatl} L. Koester, K. Knopf, and W. Waschkowski, Z. Phys. A - Atoms and Nuclei {\bf 312}, 81 (1983).

\bibitem{NBC06} P. Navratil, C. A. Bertulani and E. Caurier, Phys. Lett B {\bf 634}, 191 (2006); Phys. Rev. C {\bf 73}, 065801 (2006).

\bibitem{TUNL_A8}  D. R. Tilley {\it et al.}, Nuclear Physics A {\bf 745}, 155 (2004).

\bibitem{Halderson06} D. Halderson, Phys. Rev. C {\bf 73}, 024612 (2006).

\bibitem{Desc94} P. Descouvemont and D. Baye, Nucl. Phys. A {\bf 567}, 341 (1994).

\bibitem{FLR55} J. M. Freeman, A. M. Lane, and B. Rose, Phil. Mag. {\bf 46}, 17 (1955).

\bibitem{Greife07} U. Greife {\it et al.}, Nucl. Instrum. Methods B {\bf 261}, 1089 (2007).

\bibitem{Csoto} A. Csoto, Phys. Rev. C {\bf 61}, 024311 (2000).

\bibitem{Halderson04} D. Halderson, Phys. Rev. C {\bf 69}, 014609 (2004).

\bibitem{Volya} A. Volya, Phys. Rev. C  {\bf 79}, 044308 (2009).

\bibitem{Barker00} F. C. Barker and A. M. Mukhamedzhanov, Nucl. Phys. A {\bf 673}, 526 (2000).

\bibitem{NCSM_review} P. Navratil, S. Quaglioni, I. Stetcu and B. R. Barrett, J. Phys. G: Nucl. Part. Phys. {\bf 36}, 083101 (2009).

\bibitem{MFD} J. P. Vary, ``The Many-Fermion-Dynamics
           Shell-Model Code'', Iowa State University, 1992, unpublished.

\bibitem{Antoine} E. Caurier, G. Martinez-Pinedo, F. Nowacki, A. Poves,
                J. Retamosa and A. P. Zuker, Phys. Rev. C {\bf 59}, 2033 (1999);
	        E. Caurier and F. Nowacki, Acta Physica Polonica B
                {\bf 30}, 705 (1999).

\bibitem{AS91} F. Ajzenberg-Selove, Nucl. Phys. A {\bf 523}, 1 (1991).

\bibitem{Tra67}  W. Trachslin and L. Brown, Nucl. Phys. A {\bf 101}, 273 (1967). 

\bibitem{Hsu66} C.-C. Hsu, Y.-C. Yang, and T.-J. Lee , Chinese J. Phys. {\bf 4}, 49 (1966).

\bibitem{Ro05} C. D. Roper {\it et al.}, Phys. Rev. C {\bf 72}, 024605 (2005).

\bibitem{GUpriv} A. Galindo-Uribarri, private communication.

\bibitem{n_C12_p_C12_MCAS} G. Pisent, J. P. Svenne, L. Canton, K. Amos, S. Karataglidis, and D. van der Knijff, Phys. Rev. C {\bf 72}, 014601 (2005).

\bibitem{n_C12_MCAS} J. P. Svenne, K. Amos, S. Karataglidis, D. van der Knijff, L. Canton, and G. Pisent, Phys. Rev.  C {\bf 73}, 027601 (2006).

\bibitem{JNF09} E. D. Jurgenson, P. Navratil, and R. J. Furnstahl, Phys. Rev. Lett. {\bf 103}, 082501 (2009).

\bibitem{HPH10} G. Hagen, T. Papenbrock, and M. Hjorth-Jensen, Phys. Rev. Lett. {\bf 104}, 182501 (2010).

\bibitem{Pap10} T. Papenbrock, private communication.

\end{thebibliography}
\end{document}